\documentclass{jpp}
\usepackage{graphicx}
\usepackage{epstopdf, epsfig}
\usepackage{hyperref}
\usepackage[utf8]{inputenc}
\usepackage{bm}
\usepackage{amsmath}
\usepackage{xcolor}
\newcommand{\partder}[2]{\frac{\partial #1}{\partial #2}}
\newcommand{\der}[2]{\frac{d #1}{d #2}}
\usepackage{cleveref}
\crefname{equation}{}{}
\crefname{figure}{figure}{figures}
\usepackage{subcaption}

\title{An adjoint method for neoclassical stellarator optimization}
\author{Elizabeth J. Paul\aff{1}\corresp{ejpaul@umd.edu},Ian G. Abel\aff{1,2},Matt Landreman\aff{1},William Dorland\aff{1}}
\affiliation{
\aff{1} Institute for Research in Electronics and Applied Physics, University of Maryland, College Park, MD 20742, USA
\aff{2}Department of Physics, Chalmers University of Technology, G\"oteborg, SE-41296, Sweden}

\begin{document}

\maketitle

\begin{abstract}
    Stellarators are a promising route to steady-state fusion power. However, to achieve the required confinement, the magnetic geometry must be highly optimized.
    This optimization requires navigating high-dimensional spaces, often necessitating the use of gradient-based methods. 
    The gradient of the neoclassical fluxes is expensive to compute with classical methods, requiring $O(N)$ flux computations, where $N$ is the number of parameters.
   To reduce the cost of the gradient computation, we present an adjoint method for computing the derivatives of moments of the neoclassical distribution function for stellarator optimization. The linear adjoint method allows derivatives of quantities which depend on solutions of a linear system, such as moments of the distribution function, to be computed with respect to many parameters from the solution of only two linear systems. 
    This reduces the cost of computing the gradient to the point that the finite-collisionality neoclassical fluxes can be used within an optimization loop.
    
     With the neoclassical adjoint method, we compute solutions of the drift kinetic equation and an adjoint drift kinetic equation to obtain derivatives of neoclassical quantities with respect to geometric parameters. When the number of parameters in the derivative is large ($\mathcal{O}(10^2)$), this adjoint method provides up to a factor of 200 reduction in cost. We demonstrate adjoint-based optimization of the field strength to obtain minimal bootstrap current on a surface. With adjoint-based derivatives, we also compute the local sensitivity to magnetic perturbations on a flux surface and identify regions where tight tolerances on error fields are required for control of the bootstrap current or radial transport. Furthermore, the solve for the ambipolar electric field is accelerated using a Newton method with derivatives obtained from the adjoint method.
\end{abstract}

\section{Introduction}
The stellarator is a promising approach to magnetic confinement, as it does not require plasma current to produce rotational transform and thus is inherently steady-state and stable to current-driven modes. However, collisionless trajectories are not guaranteed to be confined in three-dimensional geometry as they are in axisymmetric systems \citep{Gibson1967,Helander2014}. This can lead to poor confinement of energetic particles and increased neoclassical transport. A possible solution is the application of numerical optimization techniques to carefully tailor the magnetic geometry for improved confinement properties. One of the first demonstrations of this technique was in the design of the Wendelstein 7-X (W7-X) stellarator \citep{Grieger1992,Lotz1991}, which was optimized for small neoclassical transport in the $1/\nu$ regime and for small bootstrap current, in addition to several other physics criteria. 

Another approach to improve confinement in stellarators is to obtain a configuration whose magnetic field strength exhibits a symmetry direction when expressed in Boozer coordinates, known as quasi-symmetry \citep{Nuhrenberg1988}. This leads to a conserved canonical momentum of the guiding center motion such that neoclassical properties are similar to those in a tokamak. However, perfect quasi-symmetry can never been achieved globally in practice \citep{Garren1991,Landreman2018a}, and it is often desirable to include symmetry-breaking components of the magnetic field strength in consideration of other design parameters, such as magneto-hydrodynamic (MHD)  stability and energetic particle confinement \citep{Nelson2003,Henneberg2019}. As one must allow for breaking of quasi-symmetry, it remains essential to include a measure of neoclassical transport such that the symmetry-breaking harmonics of the field strength do not significantly degrade the confinement. 

Neoclassical transport is governed by solutions of the drift kinetic equation, (DKE) \eqref{eq:DKE}, from which moments (e.g. radial fluxes and bootstrap current) are computed. The DKE local to a flux surface can be solved numerically \citep{Landreman2014,Belli2015}. However, this four-dimensional problem  is expensive to solve within an optimization loop, especially in low-collisionality regimes for which increased pitch-angle resolution is required to resolve the collisional boundary layer. 

Therefore, it may be desirable to consider an analytic reduction of the DKE. Under the assumption of low collisionality, a bounce-averaged DKE can be considered \citep{Beidler1995,Calvo2018}. While bounce-averaging can significantly reduce the computational cost by decreasing the spatial dimensionality, this approach typically requires restrictions on the geometry, such as closeness to omnigeneity or a model magnetic field. Additional reduction of the DKE can be made in low collisionality regimes, resulting in semi-analytic expressions. For example the effective ripple, $\epsilon_{\text{eff}}$ \citep{Nemov1999}, quantifies the geometric dependence of the $1/\nu$ radial transport and has been widely used during optimization studies \citep{Zarnstorff2001,Ku2008,Henneberg2019}. This model, though, assumes very small $E_r$, which is not always an experimentally-relevant regime. A low-collisionality semi-analytic bootstrap current model \citep{Shaing1989} is also commonly adopted for stellarator design \citep{Beidler1990,Hirshman1999}. However, this analytic expression is known to to be ill-behaved near rational surfaces. Furthermore, benchmarks with numerical solutions of the DKE in the low-collisionality limit have been shown to differ significantly from the semi-analytic model \citep{Beidler2011,Kernbichler2016}. Any analytic reduction of the DKE implies additional assumptions, such as on the collisionality, size of $E_r$, or on the magnetic geometry.


Due to the limitations of bounce-averaged and semi-analytic models, there are benefits to computing neoclassical quantities using numerical solutions to the DKE without approximation. With the numerical methods currently used for stellarator optimization, this approach becomes computationally challenging within an optimization loop. Due to their fully three-dimensional nature, optimization of stellarator geometry requires navigation through high-dimensional spaces, such as the space of the shape of the outer boundary of the plasma or the shapes of electromagnetic coils. The number of parameters required to describe these spaces, $N$, is often quite large ($\mathcal{O}(10^2)$). Knowledge of the gradient of the objective function with respect to these parameters can greatly improve the convergence to a local minimum. Once a descent direction is identified, each iteration reduces to a one-dimensional line search. Gradient-based optimization with the Levenberg-Marquardt algorithm in the STELLOPT code \citep{Strickler2004} has been widely-used in the stellarator community and led to the design of NCSX \citep{Reiman1999}.

 Although derivative information is valuable, numerically computing the derivative of a figure of merit $f$ (for example, with finite difference derivatives) can be prohibitively expensive, as $f$ must be evaluated $\mathcal{O}(N)$ times. For neoclassical optimization, this implies solving the DKE $\mathcal{O}(N)$ times; thus including finite-collisionality neoclassical quantities in the objective function is often impractical. In this work we describe an adjoint method for neoclassical optimization. With this method, the computation of the derivatives of $f$ with respect to $N$ parameters has cost comparable to solving the DKE twice, thus making the inclusion of these quantities possible within an optimization loop. In this work we obtain derivatives of neoclassical figures of merit with respect to local geometric parameters on a surface rather than the outer boundary or coil shapes. However, the geometric derivatives we compute provide an important step toward adjoint-based optimization of MHD equilibria, as discussed in section \ref{sec:equilibria_opt}.

Adjoint methods have been applied in many fields including aerodynamic engineering and computational fluid dynamics \citep{Pironneau1974,Glowinski1975}, geophysics \citep{Plessix2006,Fichtner2006}, structural engineering \citep{Allaire2005}, and tokamak divertor design \citep{Dekeyser2014a,Dekeyser2014c,Dekeyser2014b}. They have only recently been implemented for stellarator design, namely for the design of coil shapes \citep{Paul2018} and efficiently computing shape gradients for MHD equilibria \citep{Antonsen2019}. The numerical method is quite general and has the potential to greatly impact many inverse design problems in magnetic confinement fusion.  

In section \ref{sec:dke} we provide an overview of the numerical solution of the DKE local to a flux surface. In section \ref{sec:adjoint_approach} the adjoint neoclassical method is described. Two approaches to the adjoint method, termed continuous and discrete, are presented, and their implementation and benchmarks are discussed in section \ref{sec:implementation}. The adjoint method is used to compute derivatives of moments of the neoclassical distribution function with respect to local geometric quantities. The derivative information can be used to identify regions of increased sensitivity to magnetic perturbations, as discussed in section \ref{sec:local_sensitivity}. We demonstrate adjoint-based optimization in section \ref{sec:vacuum_opt} by locally modifying the field strength on a flux surface. A discussion of the application of this method for optimization of MHD equilibria is presented in \ref{sec:equilibria_opt}. 
Finally, the adjoint method is applied to accelerate the calculation of the ambipolar electric field in section \ref{sec:ambipolarity}.  

\section{Drift kinetic equation}
\label{sec:dke}

The Stellarator Fokker-Planck Iterative Neoclassical Solver (SFINCS) code \citep{Landreman2014} solves the drift kinetic equation,
\begin{equation}
\left(v_{||} \bm{b} + \bm{v}_E \right) \cdot\nabla f_{1s} - C_s(f_{1s}) = -\bm{v}_{\text{m}s} \cdot \nabla \psi \partder{f_{Ms}}{\psi},
\label{eq:DKE}
\end{equation}
for general stellarator geometry. Here $\bm{b} = \bm{B}/B$ is a unit vector in the direction of the magnetic field, $v_{||} = \bm{v}\cdot \bm{b}$ is the parallel component of the velocity, and $2\pi \psi$ is the toroidal flux. The Fokker-Planck collision operator is $C_s(f_{1s})$, linearized about a  Maxwellian $f_{Ms} = n_sv_{ts}^{-3} \pi^{-3/2} e^{-v^2/v_{ts}^2}$ where $v_{ts} = \sqrt{2T_s/m_s}$ is the thermal speed, $n_s$ is the density, $T_s$ is the temperature, $m_s$ is the mass, and the subscript indicates species. In \eqref{eq:DKE}, derivatives are performed holding $W_s = m_s v^2/2 +q_s \Phi$ and $\mu = v_{\perp}^2/2B$ fixed, where $v = \sqrt{\bm{v} \cdot \bm{v}}$ is the magnitude of velocity, $\Phi$ is the electrostatic potential, $v_{\perp} = \sqrt{v^2 - v_{||}^2}$ is the perpendicular velocity, and $q_s$ is the charge. The radial magnetic drift is
\begin{equation}
\bm{v}_{\text{m}s}\cdot \nabla \psi = \frac{m_s }{q_s B^2} \left( v_{||}^2 + \frac{v_{\perp}^2}{2} \right) \bm{b} \times \nabla B \cdot \nabla \psi,
\label{eq:radial_drift}
\end{equation}
assuming a magnetic field in MHD force balance, and $\bm{v}_E$ is the $\bm{E} \times \bm{B}$ velocity
\begin{equation}
\bm{v}_E  = \frac{\bm{B} \times \nabla \Phi}{B^2}.
\end{equation}
Throughout we assume $\Phi=\Phi(\psi)$ such that \eqref{eq:DKE} is linear. In \eqref{eq:DKE} we will not consider the effect of 
inductive electric fields, as this can be assumed to be small for stellarators without inductive current drive. We also do not consider the effects of magnetic drifts tangential to the flux surface in \eqref{eq:DKE}, as these only become important when $E_r$ is small \citep{Paul2017}. 

SFINCS solves \eqref{eq:DKE} locally on a flux surface $\psi$, thus it is four-dimensional. The SFINCS coordinates include two angles (poloidal angle $\theta$ and toroidal angle $\zeta$), speed $x_s = v/v_{ts}$, and pitch angle $\xi_s = v_{||}/v$. Specifics about the implementation of \eqref{eq:DKE} in the SFINCS code are described in appendix \ref{app:trajectory_models}. We will refer to two choices of implementation, the full trajectory model and the DKES trajectory model.  The full trajectory model maintains $\mu$ conservation as radial coupling (terms involving $\partial f_{1s}/\partial \psi$) is dropped.  While the DKES model does not conserve $\mu$ when the radial electric field $E_r \neq 0$, the adjoint operator under the DKES model takes a particularly simple form as discussed in section \ref{sec:continuous}. This model also does not introduce any unphysical constraints on the distribution function when $E_r = 0$, as occurs for the full trajectory model \citep{Landreman2014}. These constraints motivate the introduction of particle and heat sources, which are discussed in the following section. We will discuss some of the details of the implementation of the DKE in the SFINCS code as these need to be considered in arriving at the adjoint equation. However, the adjoint neoclassical approach is quite general and could be implemented in other drift kinetic codes with slight modification.

From solutions of \eqref{eq:DKE}, several neoclassical quantities are computed, including the flux surface averaged parallel flow,
\begin{align}
V_{||,s} = \frac{\left\langle B \int d^3 v \, f_{1s} v_{||} \right\rangle_{\psi}}{n_s \langle B^2 \rangle_{\psi}^{1/2}},
\label{eq:parallel_flow}
\end{align}
the radial particle flux,
\begin{align}
    \Gamma_s = \left \langle \int d^3 v \, \left(\bm{v}_{\text{m}s} \cdot \nabla \rho \right) f_{1s} \right \rangle_{\psi},
    \label{eq:particle_flux}
\end{align}
and the radial heat flux (sometimes referred to as an energy flux),
\begin{align}
    Q_s = \left \langle \int d^3 v \, \frac{m_sv^2}{2} \left(\bm{v}_{\text{m}s} \cdot \nabla \rho \right) f_{1s} \right \rangle_{\psi}.
    \label{eq:heat_flux}
\end{align}
We will also consider species-summed quantities including the bootstrap current, $J_b = \sum_s q_s n_s V_{||,s}$, the radial current, $J_r = \sum_s q_s \Gamma_s$, and the total heat flux, $Q_{\text{tot}} = \sum_s Q_s$. Here the effective normalized radius is $\rho = \sqrt{\psi/\psi_0}$, where $2\pi \psi_0$ is the toroidal flux at the boundary.

\subsection{Sources and constraints}
\label{sec:sources}
To avoid unphysical constraints on $f_{1s}$ implied by the moment equations of \eqref{eq:DKE} in the presence of a non-zero $E_r$ \citep{Landreman2014}, particle and heat sources are added to the DKE \eqref{eq:dke_model}, 
\begin{gather}
    \mathbb{L}_{0s}f_{1s} - C_s (f_{1s}) - f_{Ms} \left(x_s^2 - \frac{5}{2}\right) S_{1s}^f(\psi) - f_{Ms}\left(x_s^2-\frac{3}{2}\right) S_{2s}^f(\psi) = \mathbb{S}_{0s},
\end{gather}
where $S_{1s}^f(\psi)$ and $S_{2s}^f(\psi)$ are unknowns such that $S_{1s}^f$ provides a particle source and $S_{2s}^f$ provides a heat source. The collisionless trajectory operator in SFINCS coordinates is
\begin{gather}
    \mathbb{L}_{0s} = \dot{\bm{r}} \cdot \nabla + \dot{x}_s \partder{}{x_s} + \dot{\xi}_s \partder{}{\xi_s},
    \label{eq:L_0s}
\end{gather}
and the inhomogeneous drive term is $\mathbb{S}_{0s} = - (\bm{v}_{\text{m}s} \cdot \nabla \psi) \partial f_{Ms}/\partial \psi$. The source functions are determined via the requirement that $\langle \int d^3 v \, f_{1s} \rangle_{\psi} = 0$ and $\langle \int d^3 v \, x_s^2 f_{1s}\rangle_{\psi} = 0$ (i.e. $f_{1s}$ does not provide net density or pressure). So, the following system of equations is solved,
\begin{gather}
    \underbrace{\left[ \begin{array}{ccc}
       \mathbb{L}_{0s} -C_s  & - f_{Ms} (x_s^2-\frac{5}{2}) & -f_{Ms} (x_s^2-\frac{3}{2}) \\
       \mathbb{L}_{1s}  & 0 & 0 \\
       \mathbb{L}_{2s}  & 0 & 0 
    \end{array}
    \right]}_{\mathbb{L}_s} \underbrace{\left[
    \begin{array}{c}
    f_{1s} \\
    S_{1s}^{f} \\
    S_{2s}^{f} 
    \end{array}
    \right]}_{F_s} = \underbrace{\left[
    \begin{array}{c}
    \mathbb{S}_{0s}  \\
    0 \\
    0
    \end{array}
    \right]}_{\mathbb{S}_s}.
    \label{eq:dke_array}
\end{gather}
The velocity-space averaging operations are denoted $\mathbb{L}_{1s}f_{1s} = \langle \int d^3 v \, f_{1s} \rangle_{\psi}$ and $\mathbb{L}_{2s}f_{1s} = \langle \int d^3 v \, f_{1s} x_s^2 \rangle_{\psi}$. The full multi-species system can be written as,
\begin{gather}
\left[
\begin{array}{c}
\mathbb{L}_{1} \\
\vdots \\
\mathbb{L}_{N_{\text{species}}}
\end{array}
\right]
\left[ 
\begin{array}{c}
F_{1} \\
\vdots \\
F_{N_{\text{species}}}
\end{array}
\right] = \left[ 
\begin{array}{c}
\mathbb{S}_{1} \\
\vdots \\
\mathbb{S}_{N_{\text{species}}}
\end{array}
\right].
\label{eq:dke_species_array}
\end{gather}
Here the linear systems corresponding to each species as in \eqref{eq:dke_array} are coupled through the collision operator. We use the following notation to refer to the above system,
\begin{gather}
    \mathbb{L} F = \mathbb{S}.
    \label{eq:linear}
\end{gather}

\section{Adjoint approach}
\label{sec:adjoint_approach}

The goal of the adjoint neoclassical approach is to efficiently compute derivatives of a moment of the distribution function, $\mathcal{R}$ (e.g. $V_{||,s}, \Gamma_s, Q_s, J_b, J_r, Q_{\text{tot}})$, with respect to many parameters. Consider a set of parameters, $\Omega = \{ \Omega_i\}_{i=1}^{N_{\Omega}}$, on which $\mathcal{R}$ depends. Computing a forward difference derivative with respect to $\Omega$ requires $  N_{\Omega} + 1$ solutions of \eqref{eq:linear}. With the adjoint approach, $\partial \mathcal{R}/\partial \Omega$ can be computed with one solution of \eqref{eq:linear} and one solution of a linear adjoint equation of the same size as \eqref{eq:linear}. Thus if $N_{\Omega}$ is very large and the solution to \eqref{eq:linear} is computationally expensive to obtain, the adjoint approach can reduce the cost by $N_{\Omega}$. For stellarator optimization, it is desirable to compute derivatives with respect to parameters which describe the magnetic geometry. In fully three-dimensional geometry, $N_{\Omega}$ is $\mathcal{O}(10^2)$ and solving \eqref{eq:linear} is the most expensive part of computing $\mathcal{R}$ (rather than constructing the linear system or taking a moment of the distribution function). Thus the adjoint approach can provide a computational savings of a factor of $\mathcal{O}(10^2)$. The adjoint method is also advantageous over numerical derivatives, as it avoids additional noise from discretization error. In what follows we consider $\Omega$ to be a set of parameters describing the magnetic geometry, which will be specified in section \ref{sec:implementation}.

We compute the derivatives of $\mathcal{R}$ using two approaches. In the first approach, we define an inner product which involves integrals over the distribution function, and an adjoint operator is obtained with respect to this inner product. This we refer to as the continuous approach. In the second approach, we consider the DKE after discretization, defining an adjoint operator with respect to the Euclidean dot product. This we refer to as the discrete approach. While these approaches should provide identical results within discretization error, the advantages and drawbacks of each approach will be discussed at the end of section \ref{sec:discrete}. 

\subsection{Continuous approach}
\label{sec:continuous}
Let $F = \{F_s\}_{s=1}^{N_{\text{species}}}$ be the set of unknowns computed with SFINCS before discretization, denoted by the column vector in \eqref{eq:dke_species_array} with $F_s$ given by \eqref{eq:dke_array}. That is, $F$ consists of a set of $N_{\text{species}}$ distribution functions over $(\theta,\zeta,x_s,\xi_s)$ and their associated source functions. We define an inner product between two such quantities in the following way,
\begin{align}
\langle F,G \rangle = \sum_s \left \langle \int d^3 v \, \frac{f_{1s} g_{1s}}{f_{Ms}} \right \rangle_{\psi} + S_{1s}^f S_{1s}^g + S_{2s}^f S_{2s}^g. 
\label{eq:inner_product}
\end{align}
Here the superscript on $S_{1s}$ and $S_{2s}$ denotes the distribution function with which the source functions are associated and the sum is over species. The space of continuous functions, $F$, of this form such that $\langle F,F \rangle$ is bounded will be denoted by $\mathcal{H}$. It can be seen that \eqref{eq:inner_product} is indeed an inner product, as it satisfies  conjugate symmetry ($\langle G,F \rangle = \langle F,G \rangle$ 
$\forall F,G \in \mathcal{H}$), linearity ($\langle F + G,H \rangle = \langle F,H \rangle+ \langle G,H \rangle$ $\forall F,G,H \in \mathcal{H}$ and $\langle F, a G \rangle = a\langle F, G \rangle$ $\forall F,G \in \mathcal{H}$, $a \in \mathbb{R}$), and positive definiteness ($\langle F, F \rangle \geq 0$ and $\langle F,F \rangle = 0$ only if $F = 0$ $\forall F \in \mathcal{H}$) \citep{Rudin2006}. 
This implies that if $\mathcal{H}$ is finite-dimensional, then for any linear operator $L$ there exists a unique adjoint operator $L^{\dagger}$ such that $\langle LF,G \rangle = \langle F, L^{\dagger}G \rangle$ for all $F, G \in \mathcal{H}$. While here $\mathcal{H}$ is not finite-dimensional, we will show that such an adjoint operator exists for this inner product. 

Note that the norm associated with this inner product $|| F || = \sqrt{\langle F,F \rangle}$ is similar to the free energy norm, 
\begin{gather}
    W = \sum_s \left \langle \int d^3 v \, \frac{T_s f_{1s}^2}{2f_{Ms}} \right \rangle_{\psi},
\end{gather}
which obeys a conservation equation in gyrokinetic theory \citep{Krommes1994,Abel2013,Landreman2015}. The choice of inner product \eqref{eq:inner_product} is advantageous, as the linearized Fokker-Planck collision operator becomes self-adjoint for species linearized about Maxwellians with the same temperature. In what follows, we assume that all included species are of the same temperature. This assumption could be lifted, with a modification to the collision operator that appears in the adjoint equation (see appendix \ref{app:collision}). This assumption is not necessary when using the discrete approach (see section \ref{sec:discrete}). 

Consider a moment of the distribution function $\mathcal{R} \in \{ V_{||,s}, \Gamma_s, Q_s, J_b, J_r, Q_{\text{tot}}\}$, which can be written as an inner product with a vector $\widetilde{\mathcal{R}} \in \mathcal{H}$,
\begin{gather}
    \mathcal{R} = \langle F, \widetilde{\mathcal{R}} \rangle,
    \label{eq:inner_product_R}
\end{gather}
according to \eqref{eq:inner_product}.
For example, 
\begin{gather}
    \widetilde{J_r} = \left[ \begin{array}{c}
    q_s \bm{v}_{\text{m}s} \cdot \nabla \psi f_{Ms} \\
    0 \\
    0
    \end{array}
    \right]_{s=1}^{N_{\text{species}}},
\end{gather}
where the column structure corresponds with that in \eqref{eq:dke_array} and \eqref{eq:dke_species_array}.

We are interested in computing the derivative of $\mathcal{R}$ with respect to a set of parameters, $\Omega = \{\Omega_i\}_{i=1}^{N_{\Omega}}$. This derivative can be computed with the chain rule,
\begin{align}
\left(\partder{\mathcal{R}}{\Omega_i}\right)_{\mathbb{L}F=\mathbb{S}} = \left(\partder{\mathcal{R}}{\Omega_i}\right)_{F} + \left\langle \widetilde{\mathcal{R}}, \left(\partder{F}{\Omega_i}\right)_{\mathbb{L} F = \mathbb{S}} \right\rangle. 
\label{eq:derivative}
\end{align}
 The subscripts in \eqref{eq:derivative} denote the quantity that is held fixed while the derivative is computed. The first term on the right hand side accounts for the explicit dependence on $\Omega_i$ while the second accounts for the implicit dependence on $\Omega_i$ through $F$. Here $\left(\partial F/\partial \Omega_i\right)_{\mathbb{L} F =\mathbb{S}}$ can be computed by considering perturbations to the linear system \eqref{eq:linear}, noting that in general both $\mathbb{L}$ and $\mathbb{S}$ can depend on $\Omega$,
\begin{align}
    \mathbb{L} \left(\partder{F}{\Omega_i}\right)_{\mathbb{L}F = \mathbb{S}} = 
    \left(\partder{\mathbb{S}}{\Omega_i} - \partder{\mathbb{L}}{\Omega_i} F\right).
    \label{eq:forward}
\end{align}
Computing $\left(\partial F/\partial \Omega\right)_{\mathbb{L}F=\mathbb{S}}$ using \eqref{eq:forward} requires solving $N_{\Omega}$ linear systems of the same dimension as the DKE \eqref{eq:linear}. To avoid this additional computational cost, we instead solve an adjoint equation, 
\begin{align}
    \mathbb{L}^{\dagger}q^{\mathcal{R}}=\widetilde{\mathcal{R}}.
    \label{eq:adjoint}
\end{align}

In what follows, we show that the adjoint variable, $q^{\mathcal{R}}$, can be used to compute $\left(\partial \mathcal{R}/\partial \Omega\right)_{\mathbb{L}F = \mathbb{S}}$ without solving \eqref{eq:forward} for each $\Omega_i$. Using \eqref{eq:adjoint} with \eqref{eq:derivative}, 
\begin{align}
    \left(\partder{\mathcal{R}}{\Omega_i} \right)_{\mathbb{L}F = \mathbb{S}} = \left(\partder{\mathcal{R}}{\Omega_i} \right)_F + \left \langle \mathbb{L}^{\dagger} q^{\mathcal{R}}, \left( \partder{F}{\Omega_i}\right)_{\mathbb{L}F=\mathbb{S}} \right \rangle,
\end{align}
and applying the adjoint property, we obtain
\begin{align}
    \left(\partder{\mathcal{R}}{\Omega_i} \right)_{\mathbb{L}F = \mathbb{S}}
    &= \left(\partder{\mathcal{R}}{\Omega_i} \right)_F + \left \langle q^{\mathcal{R}}, \mathbb{L} \left(\partder{F}{\Omega_i}\right)_{\mathbb{L}F=\mathbb{S}} \right \rangle. 
\end{align}
Using \eqref{eq:forward},
\begin{gather}
   \left(\partder{\mathcal{R}}{\Omega_i} \right)_{\mathbb{L}F = \mathbb{S}}
    = \left(\partder{\mathcal{R}}{\Omega_i} \right)_F + \left \langle q^{\mathcal{R}}, \left( \partder{\mathbb{S}}{\Omega_i} - \partder{\mathbb{L}}{\Omega_i} F \right) \right \rangle.  
\label{eq:derivative_adjoint}
\end{gather}
So, \eqref{eq:derivative_adjoint} provides the same derivative information as \eqref{eq:derivative}. Thus, using (\ref{eq:derivative_adjoint}), the derivative with respect to $\Omega$ can be computed with the solution to two linear systems, (\ref{eq:linear}) and (\ref{eq:adjoint}). The partial derivatives on the right hand side of \eqref{eq:derivative_adjoint} can be computed analytically by considering the explicit geometric dependence of $\mathcal{R}$, $\mathbb{L}$, and $\mathbb{S}$. 

When $N_{\Omega}$ is large, the cost of computing $\partial \mathcal{R}/\partial \Omega$ using \eqref{eq:derivative_adjoint} is dominated not by the linear solve but by constructing $\partial \mathbb{S}/\partial \Omega$ and $\partial \mathbb{L}/\partial \Omega$ and computing the inner product. Thus the cost still scales with $N_{\Omega}$. However, we obtain a significant savings in comparison with forward difference derivatives, as shown in section \ref{sec:implementation}.

The adjoint operator for each species takes the following form,
\begin{gather}
    \mathbb{L}_s^{\dagger} = \left[ \begin{array}{c c c}
        \mathbb{L}_{0s}^{\dagger} -C_s & f_{Ms} & f_{Ms} x^2 \\
        \mathbb{L}_{1s}^{\dagger} & 0 & 0 \\
        \mathbb{L}_{2s}^{\dagger} & 0 & 0  
    \end{array} \right],
    \label{eq:L_dagger}
\end{gather}
where $\mathbb{L}_{1s}^{\dagger} = 5/2\mathbb{L}_{1s}-\mathbb{L}_{2s}$ and $\mathbb{L}_{2s}^{\dagger} = 3/2 \mathbb{L}_{1s} - \mathbb{L}_{2s}$. The same column structure is used as for the forward operator \eqref{eq:dke_species_array},  $\mathbb{L}^{\dagger} = \{ \mathbb{L}_s^{\dagger} \}_{i=1}^{N_{\text{species}}}$. The quantity $\mathbb{L}_{0s}^{\dagger}$ satisfies $\langle \int d^3 v \,  g_{1s} \mathbb{L}_{0s} f_{1s}/f_{Ms}  \rangle_{\psi} = \langle \int d^3 v \, f_{1s} \mathbb{L}_{0s}^{\dagger} g_{1s}/f_{Ms} \rangle_{\psi}$ and depends on which trajectory model is applied. The expression \eqref{eq:L_dagger} can be verified by noting that 
\begin{align}
   \langle  \mathbb{L} F, G \rangle 
   &= \sum_s \left \langle \frac{f_{1s}\left((\mathbb{L}^{\dagger}_{0s} - C_s )g_{1s} + f_{Ms} \left( S_{1s}^g + S_{2s}^g x_s^2\right)\right)}{f_{Ms}}\right \rangle_{\psi} + S_{1s}^f\mathbb{L}_{1s}^{\dagger} g_{1s} + S_{2s}^f \mathbb{L}_{2s}^{\dagger} g_{1s} \nonumber \\
   &= \langle F,\mathbb{L}^{\dagger} G \rangle.
\end{align}
For the DKES trajectories the adjoint operator is,
\begin{gather}
    \mathbb{L}_{0s}^{\dagger} = - \mathbb{L}_{0s}. 
    \label{eq:dkes_adjoint}
\end{gather}
This anti-self-adjoint property is used in obtaining the variational principle which provides bounds on neoclassical transport coefficients in the DKES code \citep{Rij1989}.
For full trajectories it is,
\begin{gather}
    \mathbb{L}_{0s}^{\dagger} = -\mathbb{L}_{0s} + \frac{q_s}{T_s} \der{\Phi}{\psi} \bm{v}_{\text{m}s} \cdot \nabla \psi.
    \label{eq:full_adjoint}
\end{gather}
The anti-self-adjoint property does not hold for this trajectory model as the $\bm{E} \times \bm{B}$ drift \eqref{eq:full_ve} is no longer divergenceless. See \cref{ap:adjoint_operators} for details on obtaining these adjoint operators. 

\subsection{Discrete approach}
\label{sec:discrete}

Next we consider the discrete adjoint approach. Let $\overrightarrow{\bm{F}}$ be the set of unknowns computed with SFINCS after discretization of $F$. The linear DKE \eqref{eq:linear} upon discretization can then be written schematically as
\begin{gather}
    \overleftrightarrow{\bm{L}} \overrightarrow{\bm{F}} = \overrightarrow{\bm{S}}. 
    \label{eq:forward_discrete}
\end{gather}
In this case, we can define an inner product as the vector dot product,
\begin{gather}
    \langle \overrightarrow{\bm{F}}, \overrightarrow{\bm{G}} \rangle = \overrightarrow{\bm{F}} \cdot \overrightarrow{\bm{G}}.
\end{gather}
In real Euclidean space, the adjoint operator, $\left(\overleftrightarrow{\bm{L}}\right)^{\dagger}$, which satisfies 
\begin{gather}
    \left \langle \overleftrightarrow{\bm{L}} \overrightarrow{\bm{F}},\overrightarrow{\bm{G}} \right \rangle = \left \langle \overrightarrow{\bm{F}},\left(\overleftrightarrow{\bm{L}}\right)^{\dagger} \overrightarrow{\bm{G}} \right \rangle
\end{gather}
is simply the transpose of the matrix, $\left(\overleftrightarrow{\bm{L}}\right)^T$. Again, the moments of the distribution function, $\mathcal{R}$ can be expressed as an inner product with a vector $\overrightarrow{\bm{R}}$,
\begin{gather}
    \mathcal{R} = \langle \overrightarrow{\bm{F}}, \overrightarrow{\bm{R}} \rangle. 
\end{gather}
Using the discrete approach, the following adjoint equation must be solved
\begin{gather}
    \left(\overleftrightarrow{\bm{L}}\right)^T \overrightarrow{\bm{q}}^{\mathcal{R}} = \overrightarrow{\bm{R}}. 
    \label{eq:adjoint_discrete}
\end{gather}
The adjoint variable, $\overrightarrow{\bm{q}}^{\mathcal{R}}$, can again be used to compute $\left(\partial \mathcal{R}/\partial \Omega_i\right)_{\overleftrightarrow{\bm{L}}\overrightarrow{\bm{F}} = \overrightarrow{\bm{S}}}$, 
\begin{gather}
    \left(\partder{\mathcal{R}}{\Omega_i} \right)_{\overleftrightarrow{\bm{L}}\overrightarrow{\bm{F}}=\overrightarrow{\bm{S}}} = \left(\partder{\mathcal{R}}{\Omega_i} \right)_{\overrightarrow{\bm{F}}} + \left \langle \overrightarrow{\bm{q}}^{\mathcal{R}}, \left( \partder{\overrightarrow{\bm{S}}}{\Omega_i} - \partder{\overleftrightarrow{\bm{L}}}{\Omega_i} \overrightarrow{\bm{F}} \right) \right \rangle. 
    \label{eq:adjoint_diagnostic_discrete}
\end{gather}
As with the continuous approach, the partial derivatives on the right hand side can be computed analytically. In this way, the derivative of $\mathcal{R}$ with respect to $\Omega$ can be computed with only two linear solves, \eqref{eq:forward_discrete} and \eqref{eq:adjoint_discrete}. 

In the SFINCS implementation, the DKE is typically solved with the preconditioned GMRES algorithm. In the continuous approach, a preconditioner matrix for both the forward and adjoint operator must be $LU$-factorized. Here the preconditioner matrix is the same as the full matrix but without cross-species or speed coupling. As the adjoint matrix is sufficiently different from the forward matrix, we do not obtain convergence when the same preconditioner is used for both problems. However, in the discrete approach, the $LU$-factorization for the preconditioner of the forward matrix can be reused for the preconditioner of the adjoint matrix (If a matrix $A$ has been factorized as $A = LU$ then $A^{T} = U^T L^T$ where $U^T$ is lower triangular and $L^T$ is upper triangular). This provides a significant reduction in memory and computational cost for the discrete approach. 

Furthermore, the discrete adjoint approach provides the exact derivatives for the discretized problem.  With this method the adjoint equation is obtained using the vector dot product and matrix transpose which can be computed without any numerical approximation. The error in the derivatives obtained by the adjoint method are therefore only limited by the tolerance to which the linear solve is performed with GMRES. On the other hand, the continuous adjoint approach relies on a continuous inner product which must ultimately be approximated numerically. Thus the continuous approach provides the exact derivatives only in the limit that the discrete approximation of the inner product exactly reproduces the continuous inner product. Therefore we expect the results of the discrete and adjoint approaches to agree within discretization error, as will be demonstrated in section \ref{sec:implementation}.

The continuous approach can be advantageous in that an adjoint equation may be prescribed independently of the discretization scheme.
Note that in the discrete approach, the adjoint operator is obtained from the matrix transpose of the discretized forward operator, which implies that the same spatial and velocity resolution parameters must be used for both the forward and adjoint solutions. In this work we will employ the same discretization parameters for both the adjoint and forward problems, but this restriction is not required for the continuous approach.

\section{Implementation and benchmarks}
\label{sec:implementation}

The adjoint method has been implemented in the SFINCS code using both the discrete and continuous approaches. The magnetic geometry is specified in Boozer coordinates \citep{Helander2014} such that the covariant form of the magnetic field is,
\begin{gather}
    \bm{B} = I(\psi) \nabla \theta + G(\psi) \nabla \zeta + K(\psi,\theta,\zeta) \nabla \psi,
    \label{eq:boozer_covariant}
\end{gather}
where $I(\psi) = \mu_0 I_T(\psi)/2\pi$ and $G(\psi) = \mu_0 I_P(\psi)/2\pi$, $I_T(\psi)$ is the toroidal current enclosed by $\psi$, and $I_P(\psi)$ is the poloidal current outside of $\psi$. The contravariant form is
\begin{gather}
    \bm{B} = \nabla \psi \times \nabla \theta - \iota(\psi) \nabla \psi \times \nabla \zeta,
    \label{eq:boozer_contravariant}
\end{gather}
where $\iota(\psi)$ is the rotational transform. The Jacobian is obtained from dotting \eqref{eq:boozer_covariant} with \eqref{eq:boozer_contravariant}, 
\begin{gather}
    \sqrt{g} = \frac{G(\psi) + \iota(\psi) I(\psi)}{B^2}.
    \label{eq:jacobian}
\end{gather}
As $K(\psi,\theta,\zeta)$ does not appear in any of the trajectory coefficients (\eqref{eq:full_trajectories} and \eqref{eq:dkes_trajectories}), in the drive term in \eqref{eq:dke_model}, or in the geometric factors used to define the moments of the distribution function (\cref{eq:parallel_flow,eq:particle_flux,eq:heat_flux}), all the geometric dependence enters through $B(\psi,\theta,\zeta)$, $G(\psi)$, $I(\psi)$, and $\iota(\psi)$. We choose to use Boozer coordinates for these computations as it reduces the number of geometric parameters that must be considered, but the neoclassical adjoint method is not limited to this choice of coordinate system.

We approximate $B$ by a truncated Fourier series,
\begin{gather}
    B = \sum_{j} B_{m_jn_j}^c \cos(m_j\theta-n_j \zeta)
    \label{eq:B_Fourier},
\end{gather}
where $j$ sums over Fourier modes $m_j \leq m_{\max}$ and $|n_j| \leq n_{\max}$ such that $n_j$ is an integer multiple of $N_P$, the number of field periods. In \eqref{eq:B_Fourier}, we have assumed stellarator symmetry such that $B(-\theta,-\zeta) = B(\theta,\zeta)$, and $N_p$ symmetry such that $B(\theta,\zeta+2\pi/N_P) = B(\theta,\zeta)$. Thus we compute derivatives with respect to the parameters $\Omega = \{B_{mn}^c, I(\psi), G(\psi), \iota(\psi)$\}. Additionally, derivatives with respect to $E_r$ are computed, which are used for efficient ambipolar solutions and computing derivatives of geometric quantities at  ambipolarity (see section \ref{sec:ambipolarity}) rather than at fixed $E_r$. 

\begin{figure}
\begin{center}
\begin{subfigure}[c]{0.422\textwidth}\includegraphics[trim=1cm 6cm 7.2cm 7cm,clip,width=1.0\textwidth]{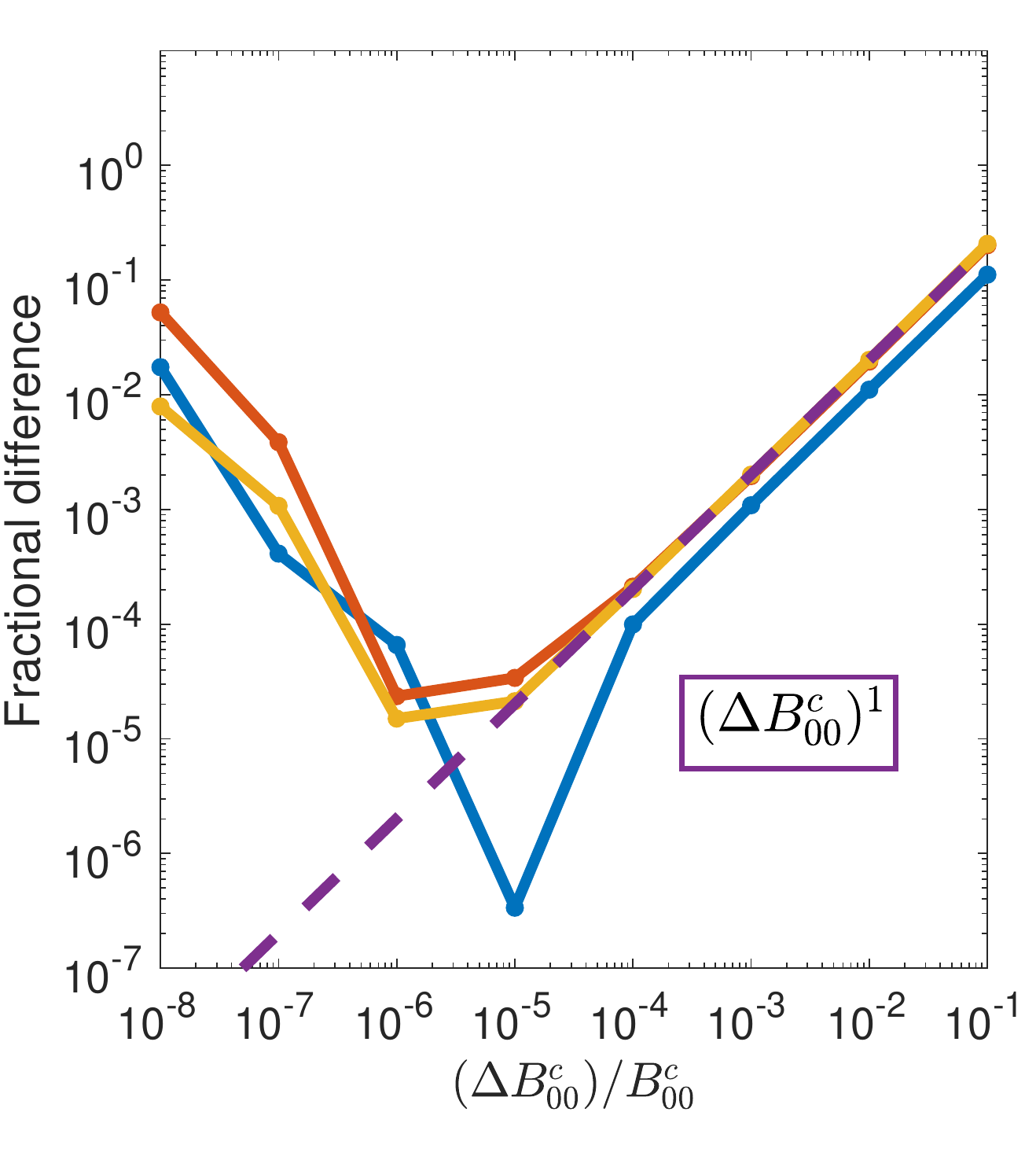}
\caption{Discrete approach}
\end{subfigure}
\begin{subfigure}[c]{0.56\textwidth}\includegraphics[trim=2cm 6cm 2cm 7cm,clip,width=1.0\textwidth]{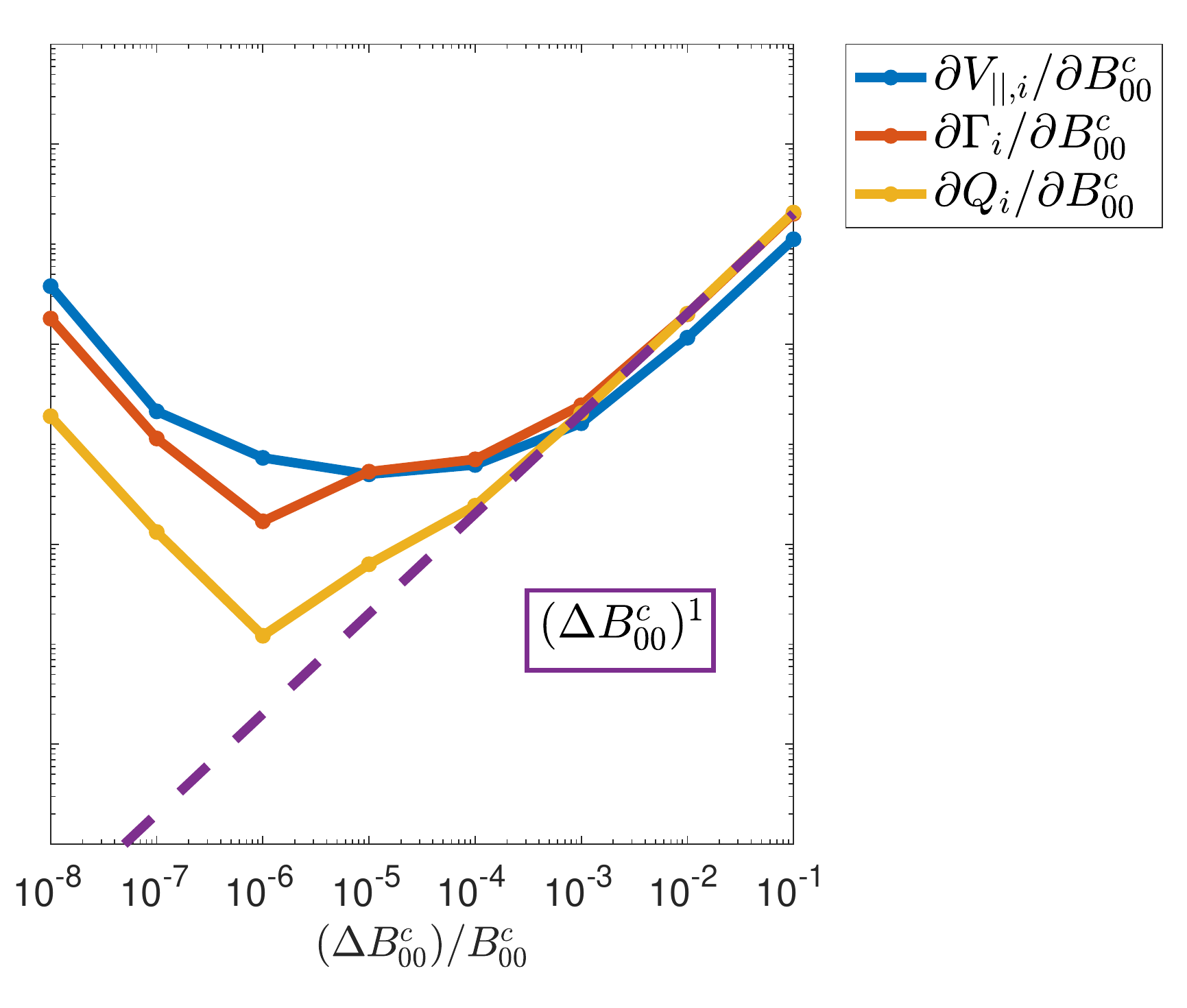}
\caption{Continuous approach}
\end{subfigure}
\caption{Fractional difference between derivatives with respect to $B_{00}^c$ computed with the adjoint method and with a forward difference derivative with step size $\Delta B_{00}^c$. The full trajectory model was used with (a) the discrete and (b) the continuous adjoint approaches.} 
\label{fig:benchmark_fixedEr}
\end{center}
\end{figure}

To demonstrate, we compute $\partial \mathcal{R}/\partial B_{00}^c$ for moments of the ion distribution function using the
discrete and continuous adjoint methods. A 3-mode model of the standard configuration W7-X geometry at $\rho = \sqrt{\psi/\psi_0} = 0.5$ is used (table 1 in \cite{Beidler2011}),
\begin{gather}
    B = B_{00}^c + B_{01}^c \cos(N_P\zeta) + B_{11}^c \cos(\theta - N_P \zeta) + B_{10}^c \cos(\theta),
\end{gather}
where $B_{01}^c = 0.04645 B_{00}^c$, $B_{11}^c = -0.04351 B_{00}^c$, and $B_{10}^c = -0.01902 B_{00}^c$.
Electron and ion ($Z=1$) species are included, and the derivatives are computed at the ambipolar $E_r$ with the full trajectory model. The derivatives are also computed with a forward difference approach with varying step size $\Delta B_{00}^c$. In figure \ref{fig:benchmark_fixedEr} we show the fractional difference between $\partial \mathcal{R}/\partial B_{00}^c$ computed using the adjoint method and with forward difference derivatives. We see that at large values of $\Delta B_{00}^c$, the adjoint and numerical derivatives begin to differ significantly due to discretization error from the forward difference approximation. The fractional error decreases proportional to $(\Delta B_{00}^c)$ as expected until the rounding error begins to dominate \citep{Sauer2012} when $\Delta B_{00}^c/(B_{00}^c)$ is approximately $10^{-4}$, where $B_{00}^c$ is the value of the unperturbed mode. The discrete and continuous approaches show qualitatively similar trends, though the minimum fractional difference is lower in the discrete approach due to the additional discretization error that arises with the continuous approach. With sufficient resolution parameters (41 $\theta$ grid points, 61 $\zeta$ grid points, 85 $\xi$ basis functions, and 7 $x$ basis functions), the fractional error of the continuous approach is $\leq 0.1 \%$ and should not be significant for most applications. We find similar agreement for other derivatives and with the DKES trajectory model.

\begin{figure}
    \centering
    \begin{subfigure}[b]{0.49\textwidth}
    \includegraphics[trim=1cm 6.5cm 2cm 7cm,clip,width=1.0\textwidth]{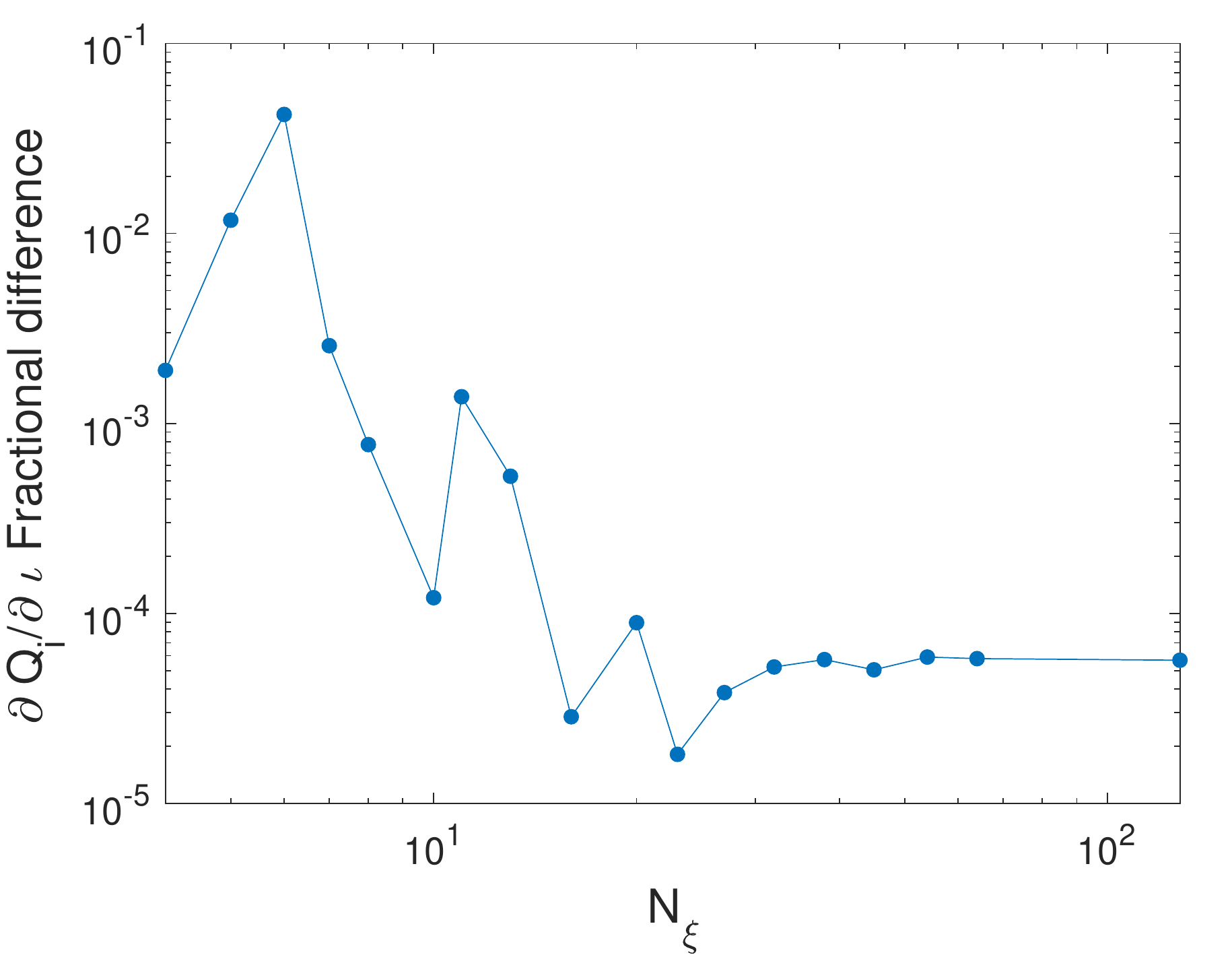}
    \caption{}
    \label{fig:continuous_discrete}
    \end{subfigure}
    \begin{subfigure}[b]{0.49\textwidth}
    \includegraphics[trim=1cm 6.5cm 2cm 7cm,clip,width=1.0\textwidth]{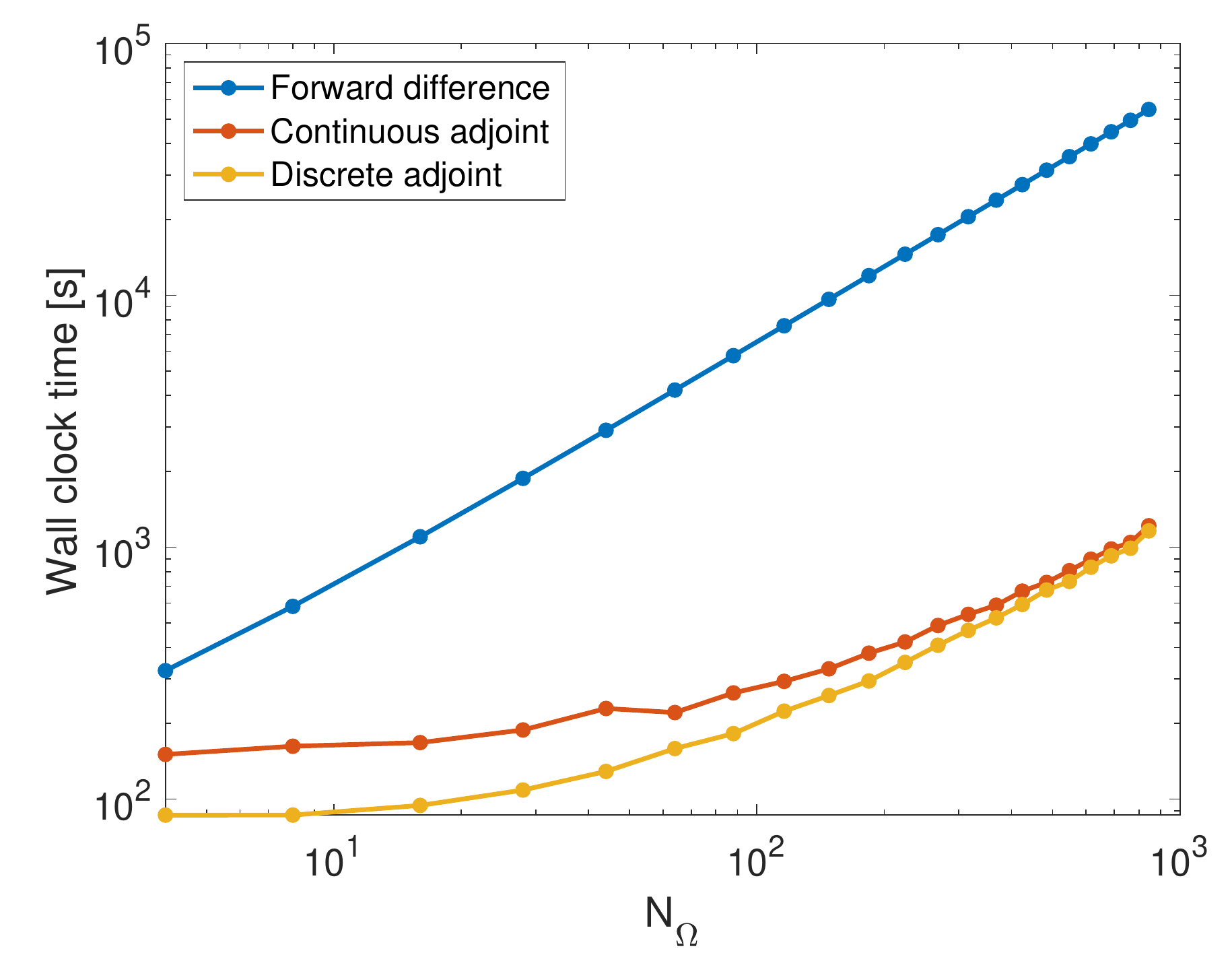}
    \caption{}
    \label{fig:computational_time}
    \end{subfigure}
    \caption{(a) The fractional difference between $\partial Q_i/\partial \iota$ computed with the continuous and discrete approaches converges with the number of pitch angle Legendre modes, $N_{\xi}$. (b) Comparison of the computational cost of computing $\partial \mathcal{R}/\partial \Omega$ with forward difference derivatives and the adjoint approach as a function of $N_{\Omega}$, the number of parameters in the gradient.}
    \label{fig:discrete_continuous_scan}
\end{figure}

To demonstrate that the discrete and continuous methods indeed produce the same derivative information, we compute the fractional difference between the derivatives computed with the two methods as a function of the resolution parameters. As an example, in figure \ref{fig:continuous_discrete} we show the fractional difference in $\partial Q_i/\partial \iota$, where $Q_i$ is the radial ion heat flux, as a function of the number of Legendre polynomials used for the pitch angle discretization, $N_{\xi}$, keeping the other resolution parameters fixed. As $N_{\xi}$ is increased, the fractional differences converges to a finite value, approximately $10^{-4}$, due to the discretization error in the other resolution parameters. Similar resolution parameters are required for the convergence of the moment itself, $Q_i$, and its derivative computed with the continuous method, $\partial Q_i/\partial \iota$. Convergence of $Q_i$ within 5\% is obtained with $N_{\xi} = 38$, similar to that required for the convergence of $\partial Q/\partial \iota$, as can be seen in figure \ref{fig:continuous_discrete}.

In figure \ref{fig:computational_time} we compare the cost of calculating derivatives of one moment with respect to $N_{\Omega}$ parameters using the continuous and discrete adjoint methods and forward difference derivatives. All computations are performed on the Edison computer at NERSC using 48 processors, and the elapsed wall time is reported. Here we include the cost of solving the linear system and computing diagnostics $N_{\Omega} + 1$ times for the forward difference approach, and the cost of solving the forward and adjoint linear systems and computing diagnostics for the adjoint approaches. The cost of the continuous approach is slightly more than that of the discrete approach due to the cost of factorizing the adjoint preconditioner. However, at large $N_{\Omega}$ the cost of computing diagnostics for the adjoint approach (e.g. computing $\partial \mathbb{S}/\partial \Omega$ and $\partial \mathbb{L}/\partial \Omega$ and performing the inner product in \eqref{eq:derivative_adjoint}) dominates that of solving the adjoint linear system; thus the discrete and continuous approaches become comparable in cost. In this regime, the adjoint approach provides speed-up by a factor of approximately $50$. 

\section{Applications of the adjoint method}
\label{sec:applications}

\subsection{Local magnetic sensitivity analysis}
\label{sec:local_sensitivity}

Using the adjoint method, it is possible to compute derivatives of a moment of the distribution function with respect to the Fourier amplitudes of the field strength, $\{ \partial \mathcal{R}/\partial B_{mn}^c\}$. Rather than consider sensitivity in Fourier space, we would like to compute the sensitivity to \textit{local} perturbations of the field strength. We now quantify the relationship between these two representations of sensitivity information.

Consider the G\^{a}teaux functional derivative \citep{Delfour2011b} of $\mathcal{R}$ with respect to $B$,
\begin{gather}
    \delta \mathcal{R}(\delta B;B(\bm{r})) = \lim_{\epsilon \rightarrow 0} \frac{\mathcal{R}(B(\bm{r}) + \epsilon \delta B(\bm{r}))-\mathcal{R}(B(\bm{r}))}{\epsilon}. 
    \label{eq:functional_derivative}
\end{gather}
Here we consider a perturbation to the field strength at fixed $I(\psi)$, $G(\psi)$, and $\iota(\psi)$. As $\delta \mathcal{R}(\delta B;B(\bm{r}))$ is a linear functional of $\delta B$, by the Riesz representation theorem \citep{Rudin2006}, $\delta \mathcal{R}$ can be expressed as an inner product with $\delta B$ and some element of the appropriate space. The function $\delta B$ is defined on a flux surface, $\psi$; thus it is sensible to express $\delta \mathcal{R}$ in the following way, 
\begin{gather}
    \delta \mathcal{R}(\delta B; B(\bm{r})) = \left \langle S_{\mathcal{R}} \delta B(\bm{r}) \right \rangle_{\psi}. 
    \label{eq:magnetic_sensitivity}
\end{gather}
Here $\delta B(\bm{r})$ describes the local perturbation to the field strength, and $\delta \mathcal{R}$ quantifies the corresponding change to the moment $\mathcal{R}$. The function $S_{\mathcal{R}}$ is analogous to the shape gradient, which quantifies the change in a figure of merit which results from a differential perturbation to a shape \citep{Landreman2018}. The shape gradient will be discussed further in section \ref{sec:equilibria_opt}.

Suppose that $B$ is stellarator symmetric and $N_P$ symmetric. If $E_r = 0$, then $S_{\mathcal{R}}$ must also possess stellarator and $N_P$ symmetry (see appendix \ref{app:symmetry}). However, when $E_r \neq 0$, $S_{\mathcal{R}}$ is no longer guaranteed to have stellarator symmetry. Nonetheless, it may be desirable to ignore the stellarator-asymmetric part of $S_{\mathcal{R}}$ if an optimized stellarator-symmetric configuration is desired. For the remainder of this work, we will make this assumption, though the analysis could be extended to consider the effect of breaking of stellarator symmetry. The quantity $S_{\mathcal{R}}$ can be approximated by a truncated Fourier series under these assumptions,
\begin{gather}
    S_{\mathcal{R}} = \sum_{k} S_{m_kn_k} \cos(m_k \theta - n_k \zeta),
\end{gather}
where $k$ sums over $m \leq m_{\max}$ and $|n| \leq n_{\max}$ such that $n$ is an integer multiple of $N_P$. The quantity $\delta B(\bm{r})$ can be written in terms of perturbations to the Fourier coefficients,
\begin{gather}
    \delta B(\bm{r}) = \sum_{j} \delta B_{m_jn_j}^c \cos(m_j \theta - n_j \zeta),
\end{gather}
where again the sum is only taken over $N_P$ symmetric modes. Now $\delta \mathcal{R}$ can be written in terms of perturbations to the Fourier coefficients,
\begin{gather}
\delta \mathcal{R} = \sum_{j} \partder{\mathcal{R}}{B_{m_jn_j}^c} \delta B_{m_jn_j}^c. 
\end{gather}
In this way, \eqref{eq:magnetic_sensitivity} can be expressed as a linear system,
\begin{gather}
    \partder{\mathcal{R}}{B_{m_jn_j}^c} = \sum_k D_{jk} S_{m_kn_k},
\end{gather}
where 
\begin{gather}
    D_{jk} = V'(\psi)^{-1} \int_{0}^{2\pi} d \theta \int_0^{2\pi} d \zeta \, \sqrt{g} \cos(m_j \theta - n_j \zeta) \cos(m_k \theta - n_k \zeta).
\end{gather}
If the same number of modes is used to discretize $\delta \mathcal{R}$ and $S_{\mathcal{R}}$, then the linear system is square.

\begin{figure}
\centering
\begin{subfigure}[b]{0.49\textwidth}
\includegraphics[trim=2cm 0cm 2cm 3cm,clip,width=1.0\textwidth]{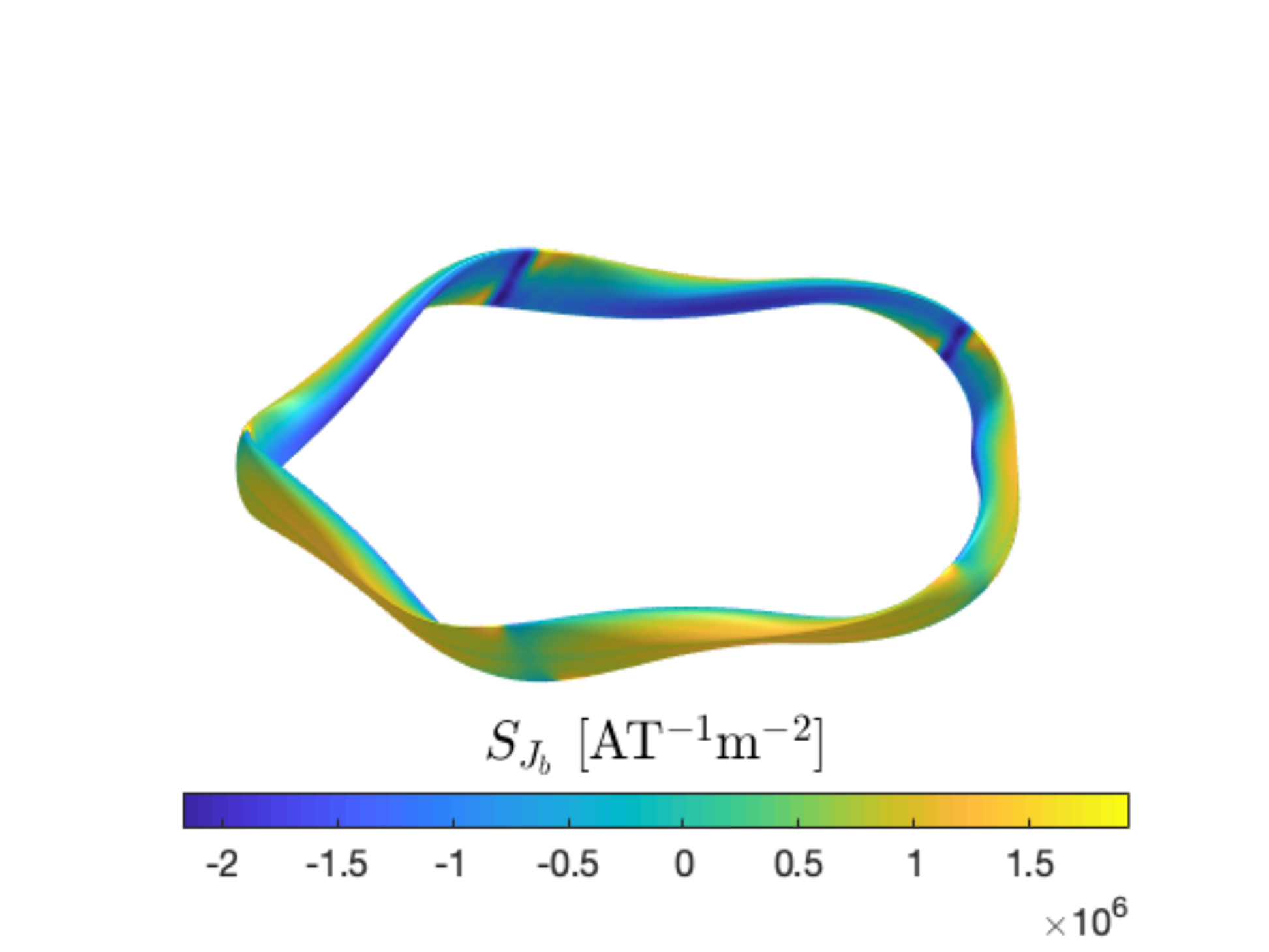}
\caption{}
\label{fig:bootstrap_local_sensitivity}
\end{subfigure}
\begin{subfigure}[b]{0.49\textwidth}
\includegraphics[trim=2cm 0cm 2cm 3cm,clip,width=1.0\textwidth]{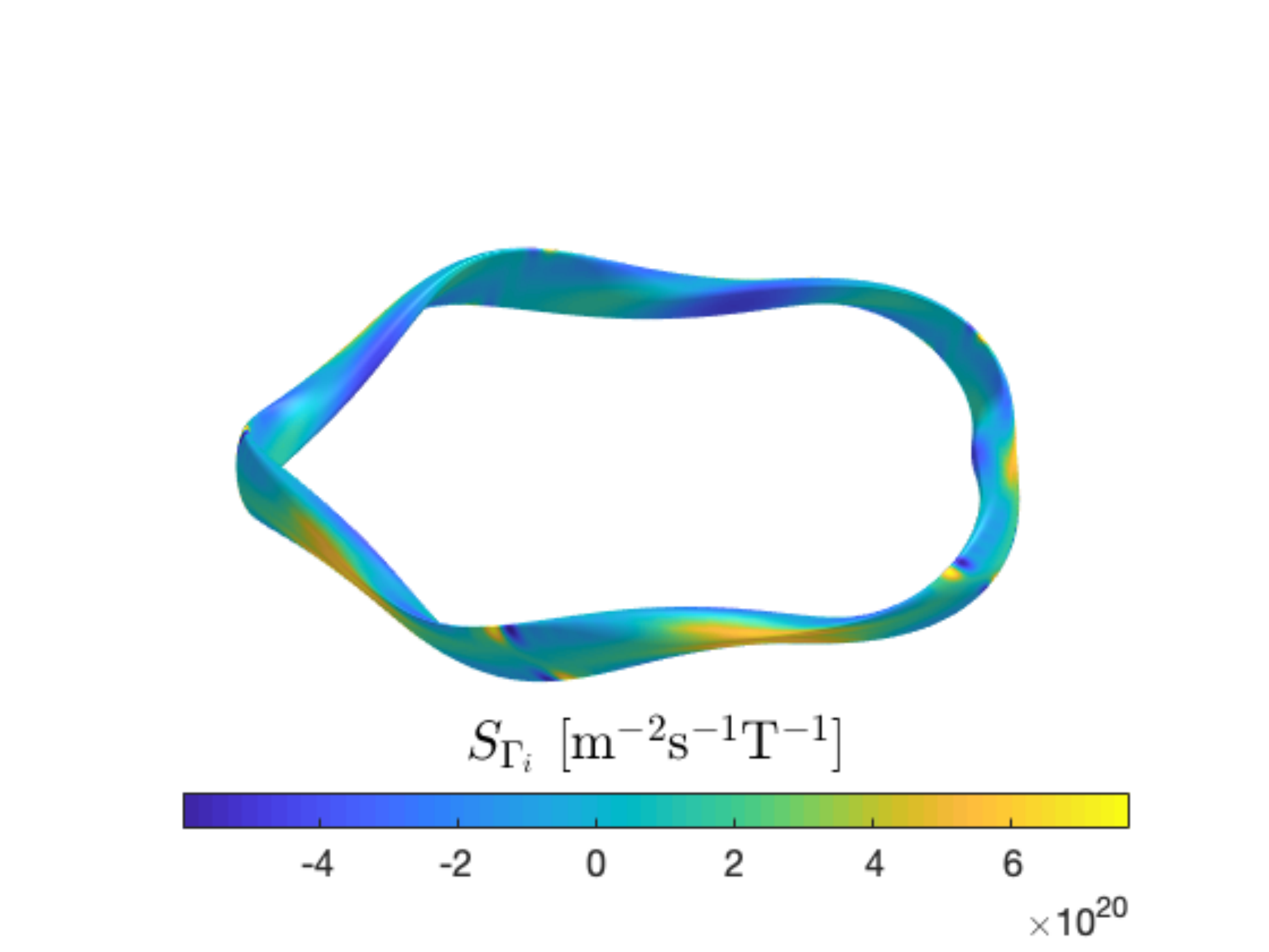}
\caption{}
\label{fig:particleFlux_sensitivity}
\end{subfigure}
\caption{(a) The local magnetic sensitivity function for the bootstrap current, $S_{J_b}$, is shown for the W7-X standard configuration. Positive values indicate that increasing the field strength at a given location will increase $J_b$ through \eqref{eq:magnetic_sensitivity}. (b) The local sensitivity function for the ion particle flux, $S_{\Gamma_i}$.}
\end{figure}

In contrast to derivatives with respect to the Fourier modes of $B$, the sensitivity function, $S_{\mathcal{R}}$, is a spatially local quantity, quantifying the change in a figure of merit resulting from a local perturbation of the field strength. In this way, $S_{\mathcal{R}}$ can inform where perturbations to the magnetic field strength can be tolerated. The sensitivity function could be related directly to a local magnetic tolerance using the method described in section 9 of \citep{Landreman2018}. In contrast with that work, here we are considering perturbations to the field strength on any flux surface rather than at the plasma boundary. However, $S_{\mathcal{R}}$ still provides insight into where trim coils should be placed or coil displacements can be tolerated without sacrificing desired neoclassical properties. The sensitivity function can also be used for gradient-based optimization in the space of the field strength on a flux surface as in section \ref{sec:vacuum_opt}.

We compute $S_{J_b}$ for the W7-X standard configuration at $\rho = 0.70$, shown in figure \ref{fig:bootstrap_local_sensitivity}. We use a fixed-boundary equilibrium that preceded the coil design and does not include coil ripple, and the full equilibrium is used rather than the truncated Fourier series considered in section \ref{sec:implementation}. The same resolution parameters are used as in section \ref{sec:implementation}, and derivatives with respect to $B_{mn}^c$ are computed for $m_{\max} = n_{\max} = 20$. The largest modes for this configuration are the helical curvature $B_{11}^c$, the toroidal curvature $B_{10}^c$, and the toroidal mirror $B_{01}^c$. We find that $S_{J_b}$ is large and negative on the inboard side, indicating that increasing the magnitude of the toroidal curvature component of $B$ would lead to an increase in $J_b$. This result is in agreement with previous analysis of the dependence of the bootstrap current on these three modes in the W7-X magnetic configuration space (\cite{Maassberg1993}), which found that at low collisionality the bootstrap current coefficients depend strongly on the toroidal curvature. In figure \ref{fig:particleFlux_sensitivity} is the sensitivity function for the ion particle flux, $S_{\Gamma_i}$, computed for the same configuration using $m_{\max} = 20$ and $n_{\max} = 25$. We find that the particle flux is more sensitive to perturbations on the outboard side in localized regions, while on the inboard side the sensitivity is relatively small in magnitude. 



\subsection{Gradient-based optimization}

\subsubsection{Optimization of field strength}
\label{sec:vacuum_opt}

As a second demonstration of the adjoint neoclassical method, we consider optimizing in the space of the field strength on a surface, taking $\Omega = \{B_{mn}^c\}$. As Boozer coordinates are used, the covariant form \eqref{eq:boozer_covariant} satisfies $(\nabla \times \bm{B}) \cdot \nabla \psi = 0$ and the contravariant form  \eqref{eq:boozer_contravariant} satisfies $\nabla \cdot \bm{B} = 0$. As we will artificially modify the field strength while keeping other geometry parameters fixed, the resulting field will not necessarily satisfy both of these conditions with both the covariant and contravariant forms. While there is no guarantee that the resulting field strength will be consistent with a global equilibrium solution, it provides insight into how local changes to the field strength can impact neoclassical properties. As a second step, the outer boundary could be optimized to match the desired field strength on a single surface. In section \ref{sec:equilibria_opt}, we discuss how the derivatives computed in this work could be coupled to optimization of an MHD equilibrium. 

\begin{figure}
    \centering
    \begin{subfigure}[b]{0.49\textwidth}
    \includegraphics[trim=1cm 6cm 2cm 6cm,clip,width=1.0\textwidth]{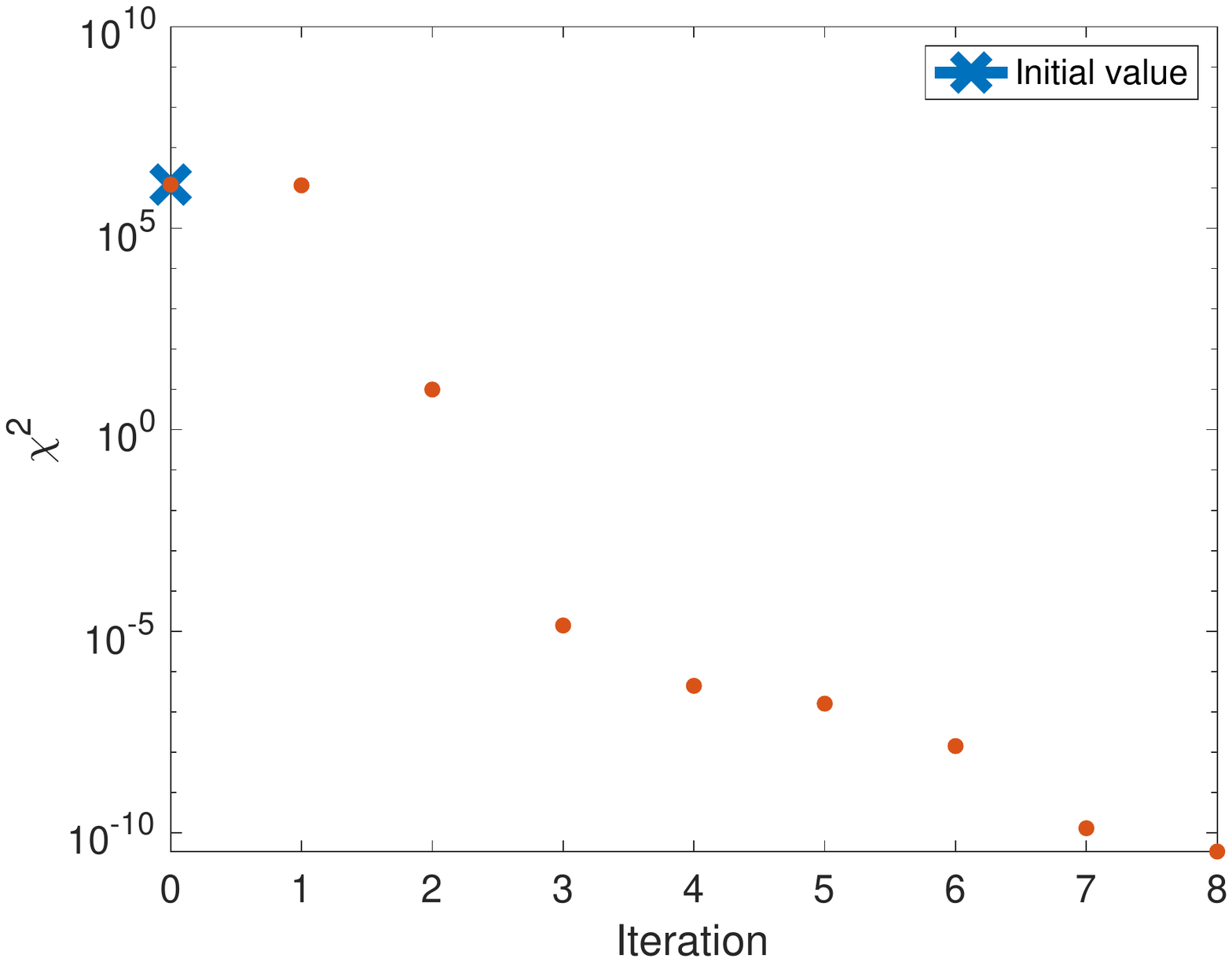}
    \caption{}
    \label{fig:bfgs_convergence}
    \end{subfigure}
    \begin{subfigure}[b]{0.49\textwidth}
    \includegraphics[trim=1cm 6cm 2cm 6cm,clip,width=1.0\textwidth]{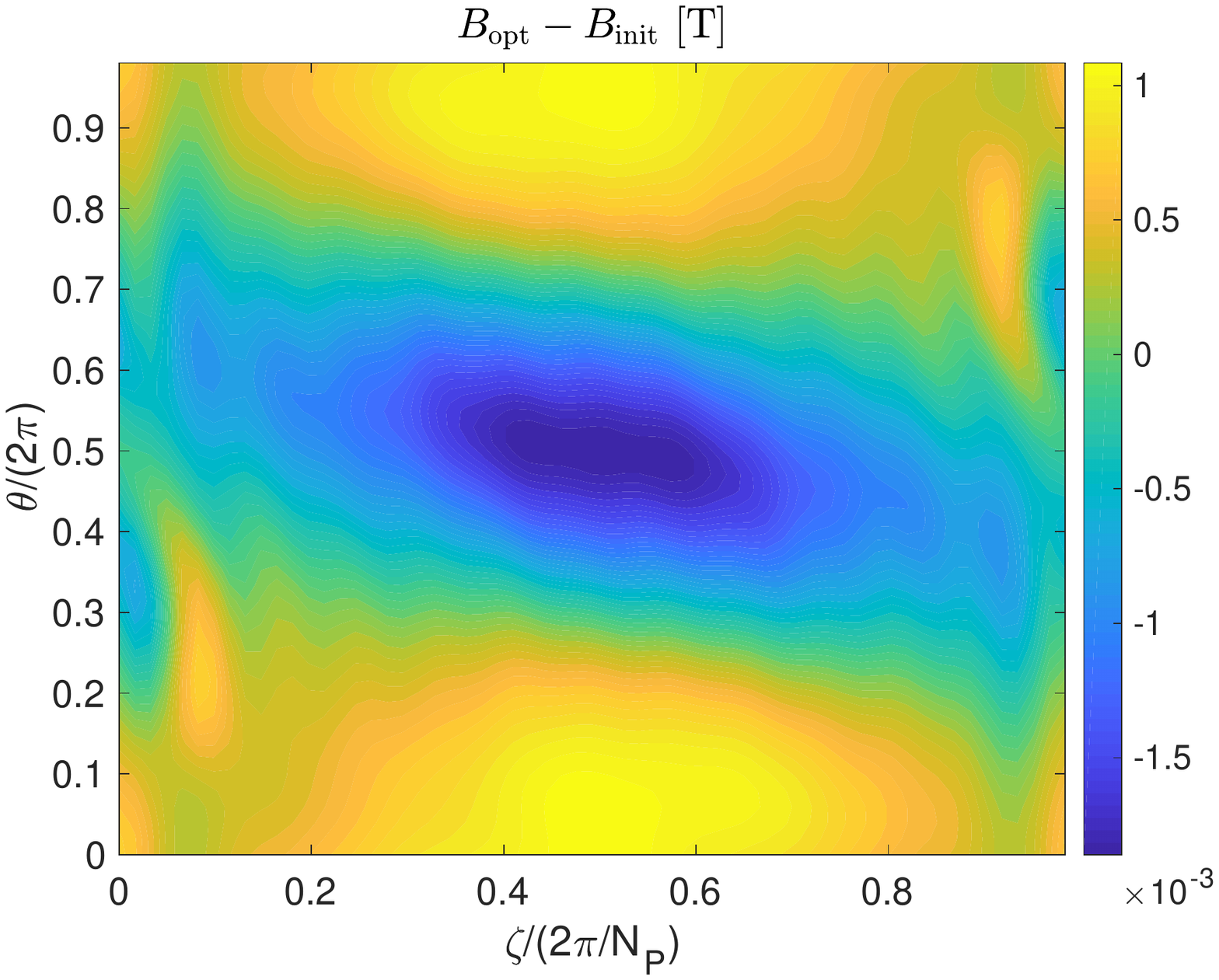}
    \caption{}
    \label{fig:B_opt}
    \end{subfigure}
    \caption{(a) Convergence of $\chi^2 = J_b^2$ for optimization over $\Omega = \{B_{mn}^c\}$ with an adjoint-based BFGS method. (b) The change in field strength from the initial to optimized configuration.}
    \label{fig:opt}
\end{figure}

We perform optimization with a BFGS quasi-Newton method \citep{Nocedal1999} using an objective function $\chi^2 = J_b^2$. A backtracking line search is used at each iteration to find a step size that satisfies a condition of sufficient decrease of $\chi^2$. We use the same equilibrium as in section \ref{sec:local_sensitivity}, retaining modes $m \leq 12$ and $|n| \leq 12$, and compute derivatives with respect to these modes. Convergence to $\chi^2 \leq 10^{-10}$ was obtained within 8 BFGS iterations (28 function evaluations), as shown in figure \ref{fig:bfgs_convergence}. The difference in field strength between the initial and  optimized configuration, $B_{\text{opt}}-B_{\text{init}}$, is shown in figure \ref{fig:B_opt}. As expected from the analysis in section \ref{sec:local_sensitivity}, the field strength increased on the outboard side and decreased on the inboard side in comparison with $B_{\text{init}}$.

\subsubsection{Optimization of MHD equilibria}
\label{sec:equilibria_opt}
The local sensitivity function, $S_{\mathcal{R}}$, along with $\partial \mathcal{R}/\partial I$, $\partial \mathcal{R}/\partial G$, and $\partial \mathcal{R}/\partial \iota$, can be used to determine how perturbations to the outer boundary of the plasma, $\partial \Gamma$, result in perturbations to $\mathcal{R}$. This is quantified through the idea of the shape gradient, which is described below. The partial derivatives of $\mathcal{R}$ can be computed with the adjoint method outlined in section \ref{sec:adjoint_approach}, and the shape gradient can be obtained with only one additional MHD equilibrium solution through the application of another adjoint method. 

Consider a figure of merit which is integrated over a toroidal domain, $\Gamma$,
\begin{gather}
    f_{\mathcal{R}}(\Gamma) = \int_{\Gamma} d^3 x \, w(\psi) \mathcal{R}(\psi),
\end{gather}
where $w(\psi)$ is a weighting function. That is, SFINCS is run on a set of $\psi$ surfaces within $\Gamma$ and the volume integral is computed numerically. Here we consider $\partial \Gamma$ to be the plasma boundary used for a fixed-boundary MHD equilibrium calculation. The perturbation to $f_{\mathcal{R}}$ resulting from normal perturbation to $\partial \Gamma$ can be written in the following form,
\begin{gather}
    \delta f_{\mathcal{R}}(\Gamma;\delta \bm{r}) = \int_{\partial \Gamma} d^3 x \, \left( \delta \bm{r} \cdot \bm{n} \right) \mathcal{G},
\end{gather}
under certain assumptions of smoothness \citep{Delfour2011a}. This can be thought of as another instance of the Riesz representation theorem, as $\delta f_{\mathcal{R}}$ is a linear functional of $\delta \bm{r}$. Here $\bm{n}$ is the outward unit normal on $\partial \Gamma$ and $\delta \bm{r}$ is a vector field describing the perturbation to the surface. Intuitively, only normal perturbations to $\partial \Gamma$ result in a change to $f_{\mathcal{R}}$. The shape gradient is $\mathcal{G}$, which quantifies the contribution of a local normal perturbation of the boundary to the change in $f_{\mathcal{R}}$. The shape gradient can be used for fixed-boundary optimization of equilibria or for analysis of sensitivity to perturbations of magnetic surfaces. It can be computed using a second adjoint method, where a perturbed MHD force balance equation is solved with the addition of a bulk force which depends on derivatives computed from the neoclassical adjoint method \citep{Antonsen2019}. While the continuous neoclassical adjoint method described in this work arises from the self-adjointness of the linearized Fokker-Planck operator, the adjoint method for MHD equilibria arises from the self-adjointness of the MHD force operator. In practice these two adjoint methods could be coupled by first computing an MHD equilibrium solution, computing neoclassical transport and its geometric derivatives from this equilibrium with the neoclassical adjoint method, and passing these derivatives back to the equilibrium code to compute the shape gradient with the perturbed MHD adjoint method. In this way derivatives of neoclassical quantities with respect to the shape of the outer boundary are computed with only two equilibrium solutions and two DKE solutions. This calculation will be reported in a future publication.

Rather than solve an additional adjoint equation, the outer boundary could be optimized by numerically computing derivatives of $\{B_{mn}^c(\psi),G(\psi),I(\psi)\}$ with respect to the double Fourier series describing the outer boundary shape in cylindrical coordinates, $\{R_{mn}^c, Z_{mn}^s\}$, using a finite difference method. This could be done using the STELLOPT code \citep{Spong1998,Reiman1999} with BOOZ\_XFORM \citep{Sanchez2000} to perform the coordinate transformation. For example, if  the rotational transform is held fixed in the VMEC equilibrium calculation \citep{Hirshman1983}, the derivative of a moment, $\mathcal{R}$, with respect to a boundary coefficient, $R_{mn}^c$, can be computed as,
\begin{gather}
    \partder{\mathcal{R}(\psi)}{R_{mn}^c} = \sum_{m'n'}\partder{\mathcal{R}(\psi)}{B_{m'n'}^c(\psi)} \partder{B_{m'n'}^c(\psi)}{R_{mn}^c} + \partder{\mathcal{R}(\psi)}{G(\psi)}\partder{G(\psi)}{R_{mn}^c} + \partder{\mathcal{R}(\psi)}{I(\psi)}\partder{I(\psi)}{R_{mn}^c},
\end{gather}
where $\partial \mathcal{R}(\psi)/\partial B_{mn}^c(\psi)$, $\partial \mathcal{R}(\psi)/\partial G(\psi)$, and $\partial \mathcal{R}(\psi)/\partial I(\psi)$ are computed with the neoclassical adjoint method and $\partial B_{mn}^c(\psi)/\partial R_{mn}^c$, $\partial G(\psi)/\partial R_{mn}^c$, and $\partial I(\psi)/\partial R_{mn}^c$ are computed with finite difference derivatives using STELLOPT. Similarly, derivatives of $\{B_{mn}^c(\psi),G(\psi),I(\psi)\}$ could be computed with respect to coil parameters using a free-boundary equilibrium solution, allowing for direct optimization of neoclassical quantities with respect to coil shapes. The neoclassical calculation with SFINCS is typically significantly more expensive than the equilibrium calculation (for the geometry discussed in section \ref{sec:local_sensitivity} fixed-boundary VMEC took 54 seconds while SFINCS took 157 seconds on 4 processors of the NERSC Edison computer). As such, combining adjoint-based with finite difference derivatives can still result in a significant computational savings. 

\subsection{Ambipolarity}
\label{sec:ambipolarity}

As stellarators are not intrinsically ambipolar, the radial electric field is not truly an independent parameter. The ambipolar $E_r$ must be obtained which satisfies the condition $J_r(E_r) = 0$. The application of adjoint-based derivatives for computing the ambipolar solution is discussed in section \ref{sec:ambipolar_sol}. An adjoint method to compute derivatives with respect to geometric parameters at fixed ambipolarity is discussed in section \ref{sec:deriv_ambipolarity}.

\subsubsection{Accelerating ambipolar solve}
\label{sec:ambipolar_sol}

 A non-linear root finding algorithm must be used to compute the ambipolar $E_r$. This root-finding can be accelerated with derivative information, such as with a Newton-Raphson method \citep{Press2007}. The derivative required, $\partial J_r/\partial E_r$, can be computed with the discrete or continuous adjoint method as described in section \ref{sec:adjoint_approach} with the replacement $\Omega_i \rightarrow E_r$, considering $\mathcal{R} = J_r$.
 
We implement three non-linear root finding methods: Brent's method \citep{Brent2013}, the Newton-Raphson method, and a hybrid between the bisection and Newton-Raphson methods \citep{Press2007}. Brent's method guarantees at least linear convergence by combining quadratic interpolation with bisection and does not require derivatives. The Newton-Raphson method can provide quadratic convergence under certain assumptions but in general is not guaranteed to converge. If an iterate lies near a stationary point or a poor initial guess is given, the method can fail. For this reason we implement the hybrid method, which combines the possible quadratic convergence properties of Newton-Raphson with the guaranteed linear convergence of the bisection method. Both Brent's method and the hybrid method require the root to be bracketed, and therefore may require additional function evaluations in order to obtain the bracket.

\begin{figure}
    \centering
    \includegraphics[trim=1cm 6cm 2cm 7.5cm,clip,width=0.49\textwidth]{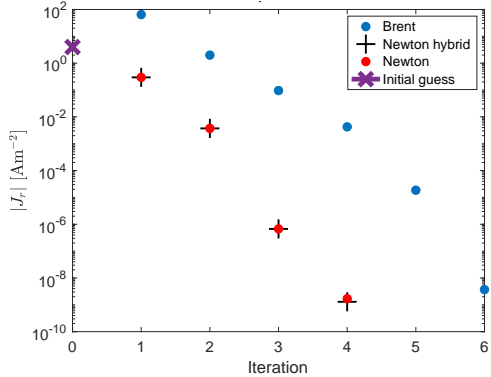}
    \caption{The ambipolar root is  obtained with Brent, Newton-Raphson, and Newton hybrid nonlinear root solvers. The derivatives obtained with the adjoint method provide better convergence properties for the Newton methods.}
    \label{fig:root_finding}
\end{figure}

We compare these methods in figure \ref{fig:root_finding}, using the W7-X standard configuration considered in section \ref{sec:local_sensitivity} with the full trajectory model and the discrete adjoint approach, beginning with an initial guess of $E_r = -10$ kV/m with bounds at $E_r^{\min} = -100$ kV/m and $E_r^{\max} =$ 100 kV/m. The root is located at $E_r =-3.56$ kV/m. For this example, the hybrid and Newton methods had nearly identical convergence properties, though the Newton method is less expensive as it does not require $J_r$ to be evaluated at the bounds of the interval. To obtain the same tolerance, the Newton method provided a 14\% savings in wall clock time over Brent's method. 

In the above discussion we have made the assumption that there is only one stable root of interest. Of course a given configuration may possess several roots, especially if the ions and electrons are in different collisionality regimes \citep{Hastings1985}. Multiple roots can be obtained by performing several root solves with different initial values and brackets, which could be trivially parallelized. Thus the adjoint method could still provide an acceleration in this more general case.

\subsubsection{Derivatives at ambipolarity}
\label{sec:deriv_ambipolarity}

The adjoint method described in section \ref{sec:adjoint_approach} assumes that $E_r$ is held constant when computing derivatives with respect to $\Omega$. However, $E_r$ cannot truly be determined independently from geometric quantities, as the ambipolar solution should be recomputed as the geometry is altered. It is therefore desirable to compute derivatives at fixed ambipolarity (fixed $J_r = 0$) rather than at fixed $E_r$. This is performed by solving an additional adjoint equation, 
\begin{gather}
    \mathbb{L}^{\dagger} q^{J_r} = \widetilde{J_r},
    \label{eq:J_r_adjoint}
\end{gather}
in the continuous approach or
\begin{gather}
    \left( \overleftrightarrow{\bm{L}} \right)^T \overrightarrow{\bm{q}}^{J_r} = \overrightarrow{\bm{J}_r},
    \label{eq:J_r_adjoint_discrete}
\end{gather}
in the discrete approach. Details are described in appendix \ref{app:ambipolar}. 

It should be noted that by computing derivatives at ambipolarity we assume that a given moment $\mathcal{R}$ is a differentiable function of the geometry at fixed $J_r = 0$. That is, this method cannot be applied to cases in which a stable root disappears as the geometry varies. As this will occur at a stationary point of $J_r(E_r)$, this situation could be avoided within an optimization loop by computing derivatives at constant $E_r$ rather than constant $J_r$ if $|\partial J_r/\partial E_r|$ falls below a given threshold at ambipolarity.

\begin{figure}
    \centering
    \begin{subfigure}[b]{0.49\textwidth}
    \includegraphics[trim=0.7cm 5.8cm 2cm 5.0cm,clip,width=1.0\textwidth]{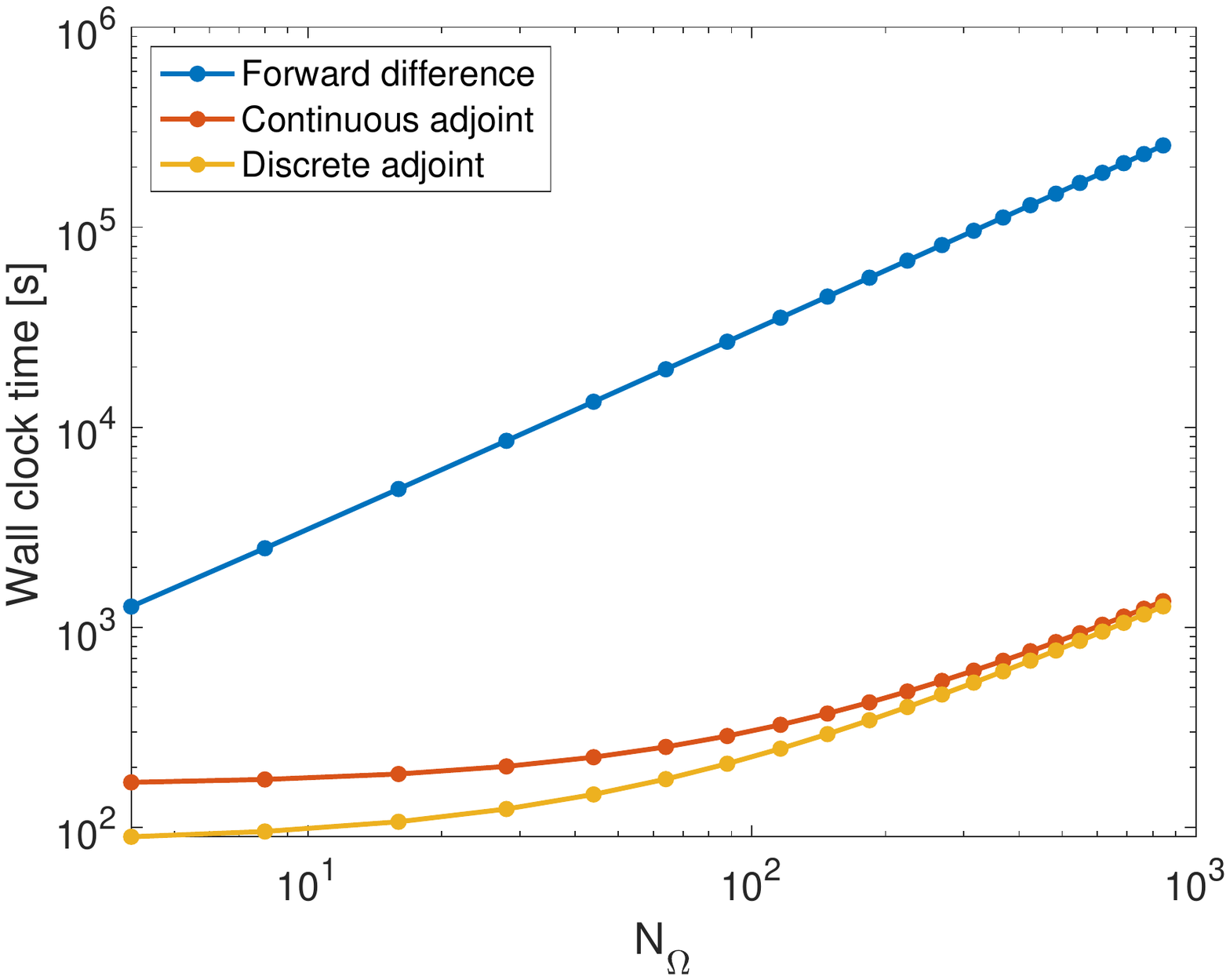}
    \caption{}
    \label{fig:cost_adjoint}
    \end{subfigure}
    \begin{subfigure}[b]{0.49\textwidth}
    \includegraphics[trim=1cm 6cm 2cm 7cm,clip,width=1.0\textwidth]{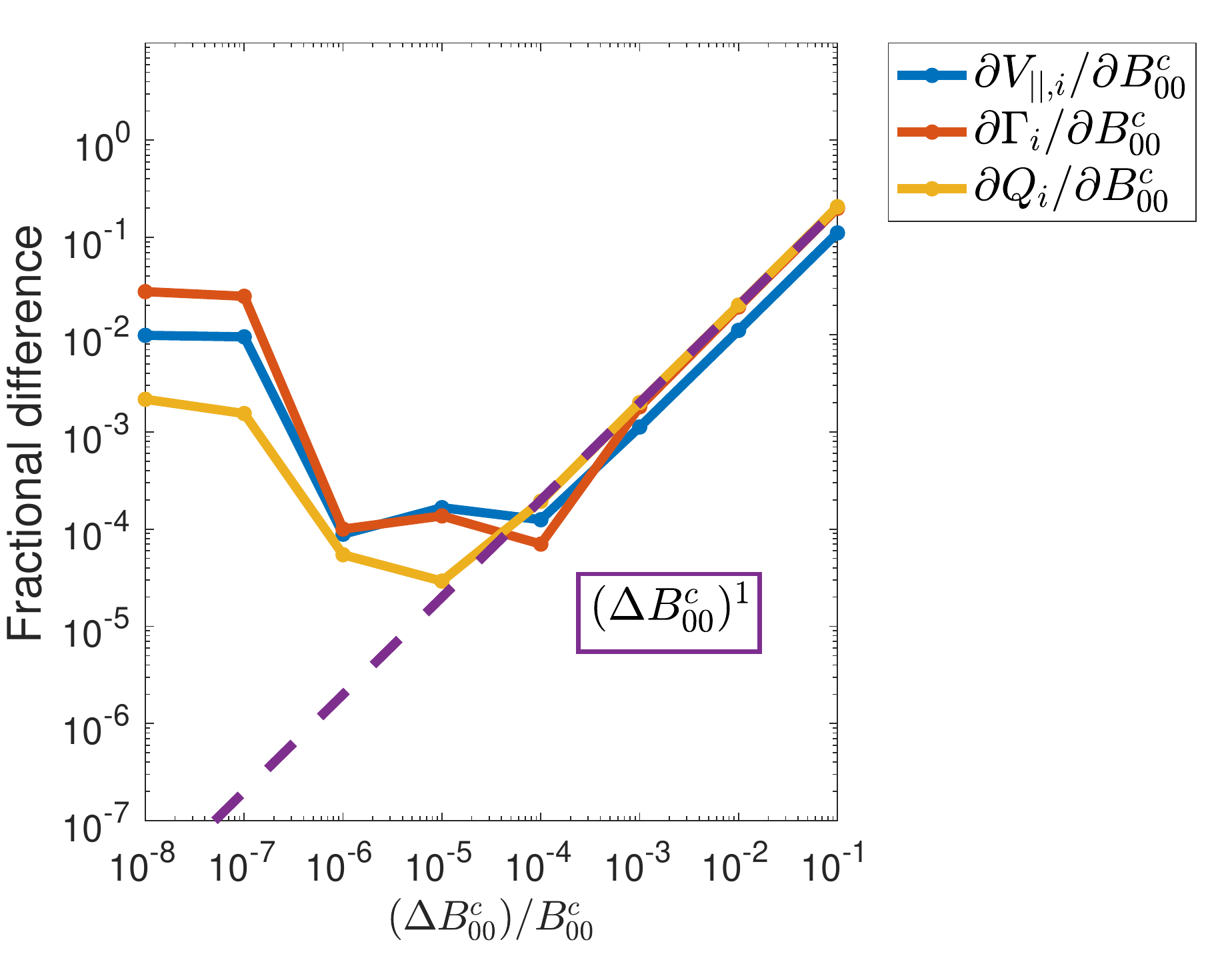}
    \caption{}
    \label{fig:ambipolar_benchmark}
    \end{subfigure}
    \caption{(a) The cost of computing the gradient $\partial \mathcal{R}/\partial \Omega$ at ambipolarity scales with $N_{\Omega}$, the number of parameters in $\Omega$. (b) The fractional difference between $\partial \mathcal{R}/\partial B_{00}^c$ at constant ambipolarity obtained with the adjoint method and with finite difference derivatives.}
\end{figure}

Although an additional adjoint solve is required, this method of computing derivatives at ambipolarity is advantageous as several linear solves are typically required to obtain the ambipolar root. A comparison of the computational cost between the adjoint method and forward difference method for derivatives at ambipolarity is shown in figure \ref{fig:cost_adjoint}. Here the full trajectory model is used, and the result for both the discrete and continuous adjoint methods are shown. For the finite difference derivative, the ambipolar solve is performed with the Brent's method at each step in $\Omega$. As in figure \ref{fig:computational_time}, we find that for large $N_{\Omega}$ the cost of the continuous and discrete approaches are essentially the same, as the cost is no longer dominated by the linear solve. When computing the derivatives at ambipolarity, both adjoint methods decrease the cost by a factor of approximately $200$ for large $N_{\Omega}$. 

In figure \ref{fig:ambipolar_benchmark} we show a benchmark between derivatives at ambipolarity, $(\partial \mathcal{R}/\partial B_{00}^c)_{J_r}$, computed with the discrete adjoint method and with forward difference derivatives. For the forward difference method, the Newton solver is used to obtain the ambipolar $E_r$ as $B_{00}^c$ is varied. As the forward difference step size $\Delta B_{00}^c$ decreases, the fractional difference again decreases proportional to $\Delta B_{00}^c)$ until it reaches a minimum when $\Delta B_{00}^c/B_{00}^c$ is approximately $10^{-4}$. In comparison with figure \ref{fig:benchmark_fixedEr}, we see that the minimum fractional difference is slightly larger at fixed ambipolarity than at fixed $E_r$, as the tolerance parameters associated with the Newton solver introduce an additional source of error to the forward difference approach.

\begin{figure}
    \centering
    \begin{subfigure}[b]{0.49\textwidth}
    \includegraphics[trim=1cm 6cm 1cm 6cm,clip,width=1.0\textwidth]{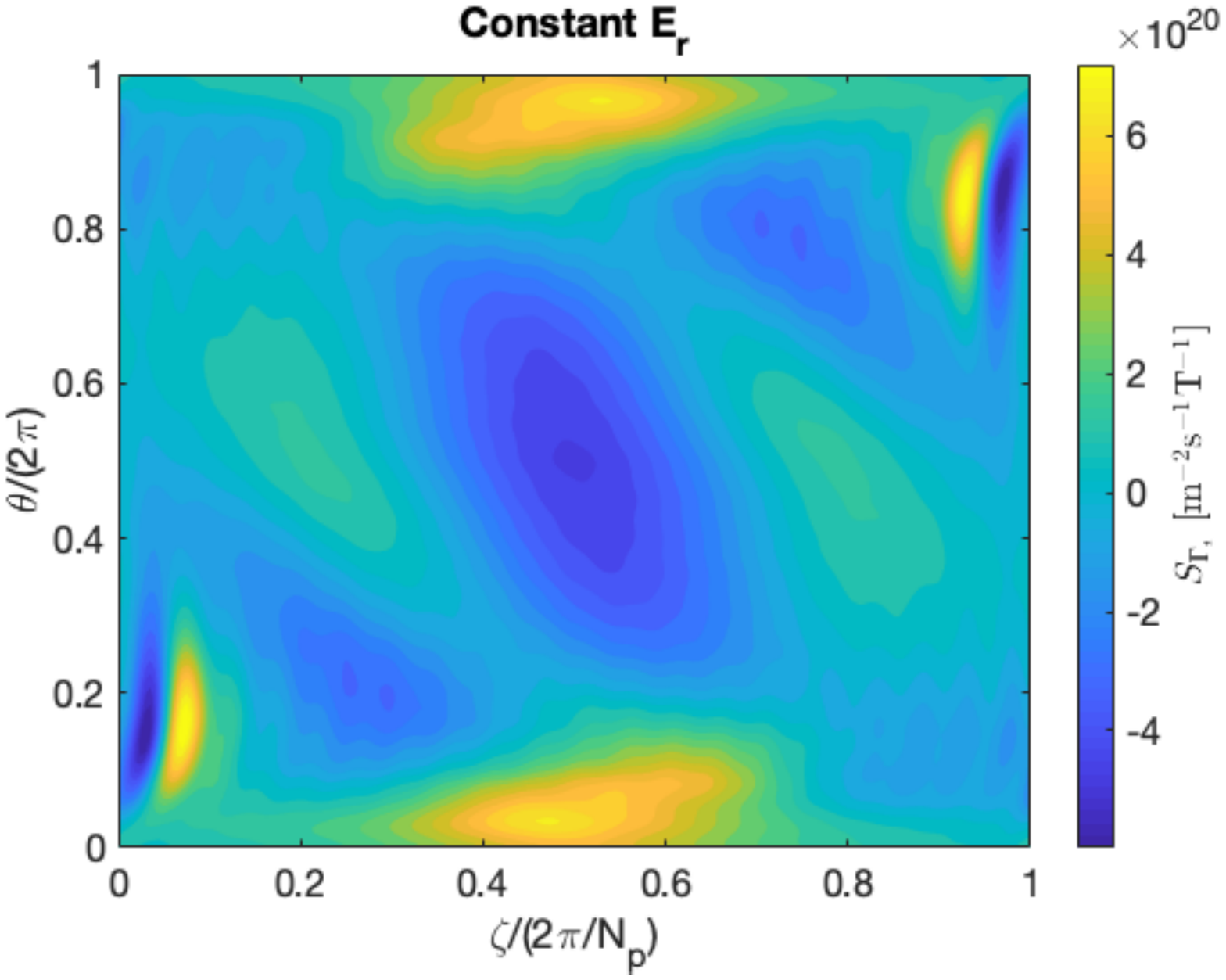}
    \caption{}
    \label{fig:S_const_Er_particle}
    \end{subfigure}
    \begin{subfigure}[b]{0.49\textwidth}
    \includegraphics[trim=1cm 6cm 1cm 6cm,clip,width=1.0\textwidth]{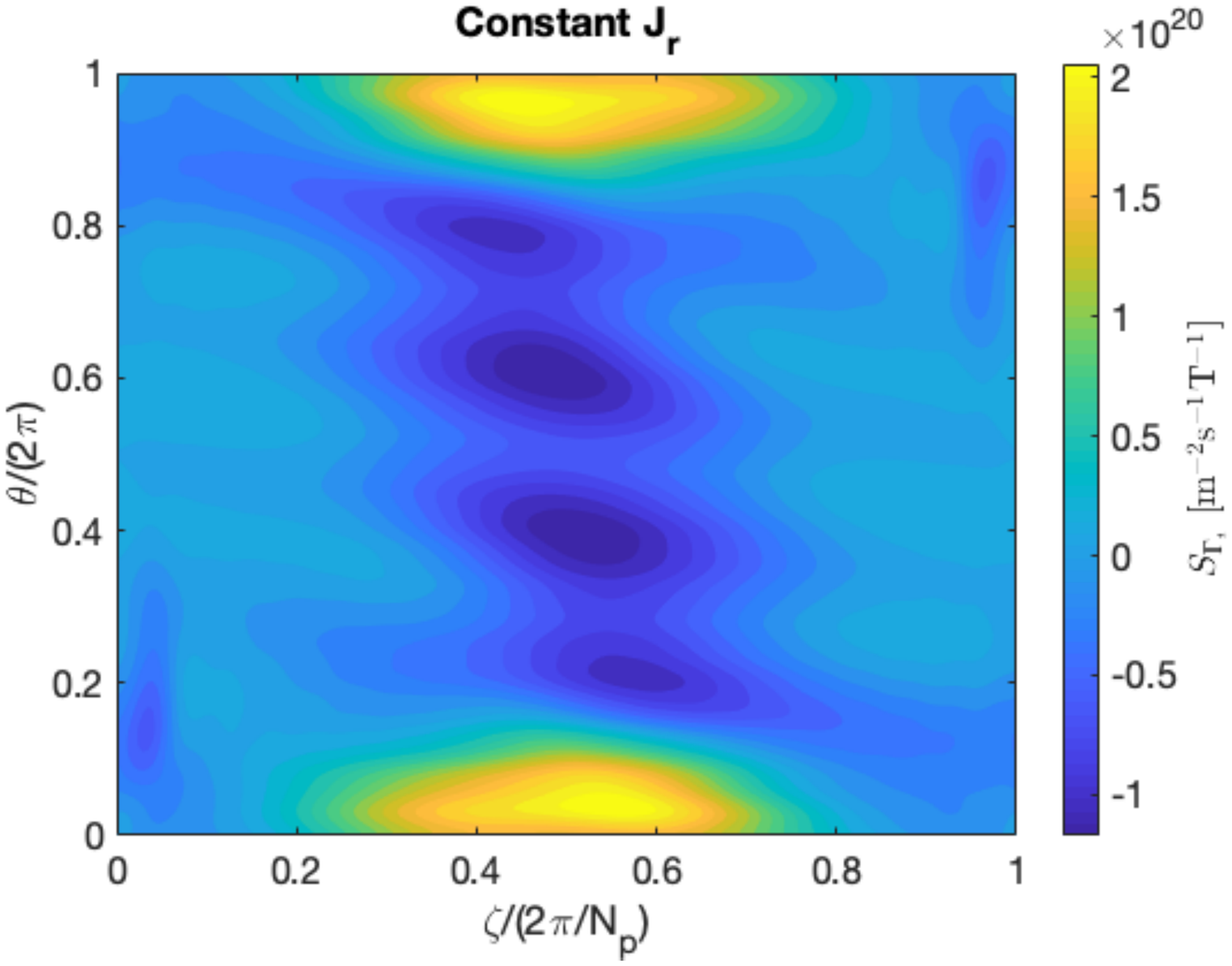}
    \caption{}
    \label{fig:S_const_Jr_particle}
    \end{subfigure}
    \begin{subfigure}[b]{0.49\textwidth}
    \includegraphics[trim=1cm 6cm 1cm 6cm,clip,width=1.0\textwidth]{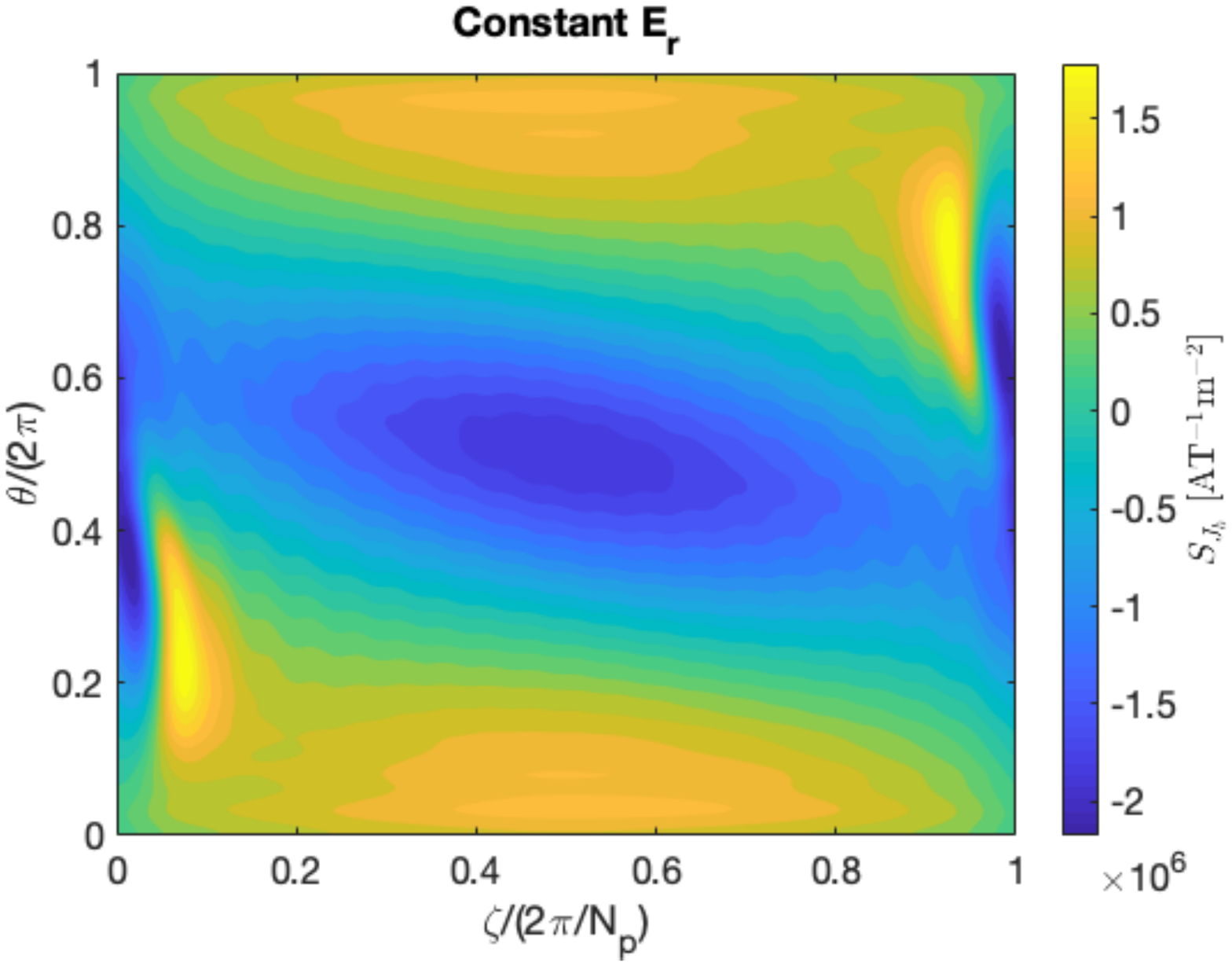}
    \caption{}
    \label{fig:S_const_Er_bootstrap}
    \end{subfigure}
    \begin{subfigure}[b]{0.49\textwidth}
    \includegraphics[trim=1cm 6cm 1cm 6cm,clip,width=1.0\textwidth]{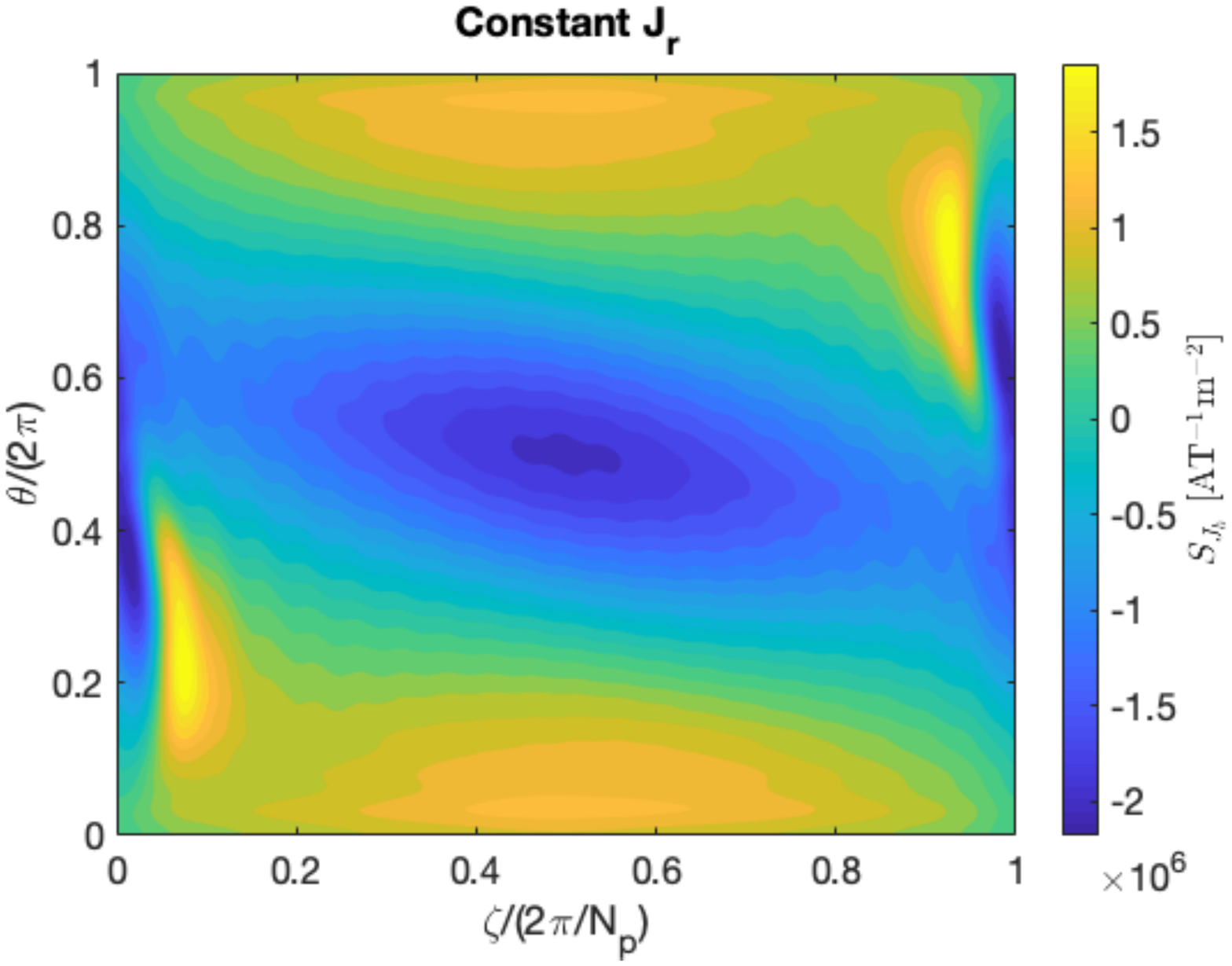}
    \caption{}
    \label{fig:S_const_Jr_bootstrap}
    \end{subfigure}
    \caption{The sensitivity function for the ion particle flux, $S_{\Gamma_i}$, is computed at (a) constant $E_r$ and (b) constant $J_r$. Similarly, $S_{J_b}$ is computed at (c) constant $E_r$ and (d) constant $J_r$.}
\end{figure}

In figures \ref{fig:S_const_Er_particle} and \ref{fig:S_const_Jr_particle} we compare the sensitivity function for the particle flux, $S_{\Gamma_i}$, computed using derivatives at constant $E_r$ with that computed at constant $J_r$. Here derivatives are computed using the discrete adjoint method with full trajectories, and the sensitivity function is constructed as described in section \ref{sec:local_sensitivity}. The configuration and numerical parameters are the same as described in section \ref{sec:local_sensitivity}. At constant $J_r$ the large region of increased sensitivity on the outboard side that appears at constant $E_r$ remains, though the overall magnitude of the sensitivity decreases. Thus it may be important to account for the effect of the ambipolar $E_r$ when optimizing for radial transport. In figures \ref{fig:S_const_Er_bootstrap} and \ref{fig:S_const_Jr_bootstrap} we perform the same comparison for $S_{J_b}$, finding the derivatives at fixed $E_r$ and at fixed $J_r$ to be virtually identical. This is to be expected, as numerical calculations of neoclassical transport coefficients for W7-X have found that the bootstrap coefficients are much less sensitive to $E_r$ than those for the radial transport (figures 18 and 26 in \cite{Beidler2011}). Furthermore, the bootstrap current in the $1/\nu$ regime is independent of $E_r$, and the finite-collisionality correction is small for optimized stellarators, such as W7-X \citep{Helander2017}. Therefore, the ambipolarity corrections to the derivatives are less important for $J_b$ than for the radial transport.

\section{Conclusions}
We have described a method by which moments $\mathcal{R}$ of the neoclassical distribution function can be differentiated efficiently with respect to many parameters. The adjoint approach requires defining an inner product from which the adjoint operator is obtained. We consider two choices for this inner product. One choice corresponds with computing the adjoint of the linear operator after discretization, and the other corresponds with computing it before discretization. In the case of the former, the Euclidean dot product can be used, and in the case of the latter, an inner product whose corresponding norm is similar to the free energy norm \eqref{eq:inner_product} is defined. In section \ref{sec:implementation}, we show that these approaches provide the same derivative information within discretization error, as expected. Both methods provide reduction in computational cost by a factor of approximately $50$ in comparison with forward difference derivatives when differentiating with respect to many ($\mathcal{O}(10^2)$) parameters. In section \ref{sec:deriv_ambipolarity} the adjoint method is extended to compute derivatives at ambipolarity. This method provides a reduction in cost by a factor of approximately $200$ over a forward difference approach. We have implemented this method in the SFINCS code, 
and similar methods
could be applied to other drift kinetic solvers.

In this work we consider derivatives with respect to geometric quantities that enter the DKE through Boozer coordinates. However, the adjoint neoclassical method we have described is much more general, allowing for many possible applications. For example, derivatives of the radial fluxes with respect to the temperature and density profiles could be used to accelerate the solution of the transport equations using a Newton method \citep{Barnes2010}. The transport solution could furthermore be incorporated into the optimization loop to self-consistently evolve the macroscopic profiles in the presence of neoclassical fluxes. Rather than simply optimizing for minimal fluxes, an objective function such as the total fusion power could be considered \citep{Highcock2018}, with optimization accelerated by adjoint-based derivatives.

Another application of the continuous adjoint formulation is correction of discretization error. The same solution obtained in section \ref{sec:continuous} can be used to quantify and correct for the error in a moment, $\mathcal{R}$, providing similar accuracy to that computed with a higher-order stencil or finer mesh without the associated cost. This method has been applied in the field of computational fluid dynamics by solving adjoint Euler equations \citep{Venditti1999,Pierce2004} and could prove useful for efficiently obtaining solutions of the DKE in low collisionality regimes.

In section \ref{sec:vacuum_opt} we have shown an example of adjoint-based neoclassical optimization, where the optimization space is taken to be the Fourier modes of the field strength on a surface, $\{B_{mn}^c\}$. While optimization within this space is not necessarily consistent with a global equilibrium solution, it demonstrates the adjoint neoclassical method for efficient optimization. In section \ref{sec:equilibria_opt}, two approaches to self-consistently optimize MHD equilibria are discussed. Further discussion and demonstration of these approaches will be provided in a future publication.

In appendix \ref{app:symmetry} we show that when $E_r = 0$ and the unperturbed geometry is stellarator symmetric, the sensitivity functions for moments of the distribution function are also stellarator symmetric. However, when $E_r \neq 0$ this is no longer true. This implies that obtaining minimal neoclassical transport in the $\sqrt{\nu}$ regime may require breaking of stellarator symmetry. In this work we have ignored the effects of stellarator symmetry-breaking, though we hope to extend this work to study these effects in the future.

\section*{Acknowledgements}
The authors acknowledge helpful discussion with L.-M. Imbert-G\'{e}rard and assistance with the STELLOPT code from S. Lazerson. This work was supported by the US Department of Energy through grants DE-FG02-93ER-54197 and DE-FC02-08ER-54964. The computations presented in this paper have used resources at the National Energy Research Scientific Computing Center (NERSC). 
Support for IGA for the initial work was provided by the Chalmers University of Technology, under the auspices of the Framework grant for Strategic Energy Research (Dnr. 2014-5392) from Vetenskapsr{\aa}adet.

\appendix

\section{Trajectory models}
\label{app:trajectory_models}

In the SFINCS coordinate system, the DKE can be written in the following way,
\begin{equation}
\dot{\bm{r}} \cdot \nabla f_{1s} + \dot{x}_s \partder{f_{1s}}{x_s} + \dot{\xi}_s \partder{f_{1s}}{\xi_s} - C_s(f_{1s}) = - \left( \bm{v}_{\text{m}s} \cdot \nabla \psi \right) \partder{f_{Ms}}{\psi}.
\label{eq:dke_model}
\end{equation}
To obtain the trajectory coefficients ($\dot{\bm{r}}$, $\dot{x}_s$, and $\dot{\xi}_s$) several approximations are made. For example, any terms that require radial coupling ($\psi$ derivatives of $f_{1s}$) cannot be retained, as this would necessitate solving a five-dimensional system.

Under the full trajectory model, the trajectory coefficients are chosen such that $\mu$ conservation is maintained as radial coupling is dropped,
\begin{align}
 \dot{\bm{r}} &= v_{||} \bm{b} + \frac{1}{B^2}\der{\Phi}{\psi} \bm{B} \times \nabla \psi \nonumber \\
 \dot{x}_s &= - \left( \bm{v}_{\text{m}s} \cdot \nabla \psi \right) \frac{q_s}{2T_s x_s} \der{\Phi}{\psi} \nonumber \\
 \dot{\xi}_s &= - \frac{1-\xi_s^2}{2B\xi_s} v_{||} \bm{b} \cdot \nabla B + \xi_s(1-\xi_s^2) \frac{1}{2B^3} \der{\Phi}{\psi} \bm{B}\times \nabla \psi \cdot \nabla B.
 \label{eq:full_trajectories}
\end{align}
Under the DKES trajectory model, the $\bm{E} \times \bm{B}$ velocity is taken to be divergenceless,
\begin{gather}
    \bm{v}_E^{\text{DKES}} = \frac{\bm{B} \times \nabla \Phi}{\langle B^2 \rangle_{\psi}}.
\end{gather}
where the flux surface average of a quantity $A$ is,
\begin{gather}
 \langle A \rangle_{\psi} = V'(\psi)^{-1}\int_0^{2\pi}d \theta \int_0^{2\pi} d \zeta \, \sqrt{g} A,
\end{gather}
$\sqrt{g} = \left( \nabla \psi \times \nabla \theta \cdot \nabla \zeta \right)^{-1}$ is the Jacobian, and $V'(\psi) = \int_0^{2\pi} d \theta \int_0^{2\pi} d \zeta \, \sqrt{g}$. Under the DKES trajectory model, the trajectory coefficients are taken to be, 
\begin{align}
\dot{\bm{r}} &= v_{||} \bm{b} + \frac{1}{\langle B^2\rangle_{\psi}} \der{\Phi}{\psi} \bm{B} \times \nabla \psi \nonumber \\
\dot{x}_s &= 0 \nonumber \\
\dot{\xi}_s &= - \frac{1-\xi_s^2}{2 B\xi_s} v_{||} \bm{b} \cdot \nabla B. 
\label{eq:dkes_trajectories}
\end{align}
These effective trajectories are adopted in the widely-used DKES code \citep{Hirshman1986,Rij1989}.

\section{Adjoint collision operator}
\label{app:collision}

We want to find an adjoint collision operator, $C_s^{\dagger}$, that satisfies the following relation,
\begin{gather}
    \left \langle \int d^3 v \, \frac{g_{1s} C_s(f_{1s})}{f_{Ms}} \right \rangle_{\psi} = \left \langle \int d^3 v\, \frac{f_{1s} C_s^{\dagger}(g_{1s})}{f_{Ms}} \right \rangle_{\psi}.
\end{gather}

The linearized Fokker-Planck collision operator can be written as
\begin{gather}
    C_s(f_{1s}) = \sum_{s'} C_{ss'}^L(f_{1s},f_{1s'}) = \sum_{s'} C_{ss'}(f_{1s},f_{Ms'}) + C_{ss'}(f_{Ms},f_{1s'}), 
    \label{eq:C_f_1s}
\end{gather}
where $s'$ sums over species. The first term on the right hand side of \eqref{eq:C_f_1s} is referred to as the test-particle collision operator, $C^T_{ss'}(f_{1s}) = C_{ss'}(f_{1s},f_{Ms'})$, and the second the field-particle collision operator, $C^F_{ss'}(f_{1s'}) = C_{ss'}(f_{Ms},f_{1s'})$. The test and field terms satisfy the following relations \citep{Rosenbluth1972,Sugama2009},
\begin{align}
    \int d^3 v \, \frac{g_{1s} C_{ss'}(f_{1s}, f_{Ms'})}{f_{Ms}} &= \int d^3 v \, \frac{f_{1s} C_{ss'}(g_{1s},f_{Ms'})}{f_{Ms}} \\
    \int d^3 v \, \frac{g_{1s} C_{ss'}(f_{Ms},f_{1s'})}{f_{Ms}}  &= \frac{T_{s'}}{T_{s}} \int d^3 v \, \frac{f_{1s'}C_{s's}(f_{Ms'},g_{1s})}{f_{Ms'}}.
\end{align}
For collisions between species of the same temperature, we see that $C_{s}(f_{1s})$ is self-adjoint. The adjoint operator with respect to the inner product \eqref{eq:inner_product} is thus,
\begin{align}
  C_s^{\dagger} &= C_s^T + \sum_{s'} \frac{f_{Ms}}{f_{Ms'}} \frac{T_{s'}}{T_s} C_{s's}^F .
\end{align}

\section{Adjoint collisionless trajectories}
\label{ap:adjoint_operators}

We want to find an adjoint operator, $\mathbb{L}_{0s}^{\dagger}$, that satisfies,
\begin{gather}
    \left \langle \int d^3 v \, \frac{g_{1s}\mathbb{L}_{0s} f_{1s}}{f_{Ms}} \right \rangle_{\psi} = \left \langle \int d^3 v \, \frac{f_{1s}\mathbb{L}_{0s}^{\dagger} g_{1s} }{f_{Ms}} \right \rangle_{\psi},
    \label{eq:L_0s_adjoint}
\end{gather}
for both trajectory models, where $\mathbb{L}_{0s}$ is defined in \eqref{eq:L_0s} with \eqref{eq:dkes_trajectories} for the DKES trajectories model and \eqref{eq:full_trajectories} for the full trajectory model. Throughout we use the velocity space element in SFINCS coordinates, $d^3 v = 2\pi v_{ts}^3 x_s^2 d \xi_s d x_s$.  

\subsection{DKES trajectories}

The operator under consideration is,
\begin{gather}
    \mathbb{L}_{0s} = v_{||} \bm{b} \cdot \nabla + \bm{v}_E^{\text{DKES}} \cdot \nabla - \frac{1-\xi_s^2}{2B\xi_s} v_{||} \bm{b} \cdot \nabla B \partder{}{\xi_s}. \label{eq:L_0s_dkes}
\end{gather}
Considering the contribution of the streaming term in \eqref{eq:L_0s_dkes} to the left hand side of \eqref{eq:L_0s_adjoint} we obtain,
\begin{gather}
    \left \langle \int d^3 v \, \frac{g_{1s} v_{||} \bm{b} \cdot \nabla f_{1s}}{f_{Ms}} \right \rangle_{\psi}
    = - \left \langle \int d^3 v \, \frac{f_{1s} v_{||} \bm{B} \cdot \nabla \left(g_{1s}/B\right)}{f_{Ms}} \right \rangle_{\psi}. \label{eq:dkes_parallel_streaming}
\end{gather}
Here the identity $\langle \nabla \cdot \bm{Q}\rangle_{\psi} = 1/V'(\psi) \partial/\partial \psi \left( V'(\psi) \langle \bm{Q} \cdot \nabla \psi \rangle_{\psi} \right)$ for any vector $\bm{Q}$ has been used. We next consider the contribution of the $\bm{E} \times \bm{B}$ drift term in \eqref{eq:L_0s_dkes},
\begin{align}
    \left \langle \int d^3 v \, \frac{g_{1s} \bm{v}_E^{\text{DKES}} \cdot \nabla f_{1s} }{f_{Ms}} \right \rangle_{\psi} 
    &= -\left \langle \int d^3 v \, \frac{f_{1s} \bm{v}_E^{\text{DKES}} \cdot \nabla g_{1s} }{f_{Ms}} \right \rangle_{\psi}.
    \label{eq:vE_dkes}
\end{align}
Here we have used the identity 
\begin{gather}
    \left \langle \bm{B} \times \nabla \psi \cdot \nabla w \right \rangle_{\psi} = 0,
    \label{eq:B_times_nabla_psi}
\end{gather}
    for any $w$. We consider the contribution of the mirror-force term in \eqref{eq:L_0s_dkes}, 
\begin{align}
 \left \langle \int d^3 v \, \frac{g_{1s}  \dot{\xi}_s}{f_{Ms}} \partder{ f_{1s}}{\xi_s} \right \rangle_{\psi} 
 &= -\left \langle \int d^3 v \, \frac{f_{1s} \dot{\xi}_s}{f_{Ms}} \partder{g_{1s}}{\xi_s} \right \rangle_{\psi} -\left \langle \int d^3 v \, \frac{v_{||}}{B} \bm{b} \cdot \nabla B \frac{g_{1s} f_{1s}}{f_{Ms}} \right \rangle_{\psi}. 
 \label{eq:dkes_mirror}
\end{align}
Combining (\ref{eq:dkes_parallel_streaming}-\ref{eq:dkes_mirror}), we obtain
\begin{gather}
    \left \langle \int d^3 v \, \frac{g_{1s} \mathbb{L}_{0s} f_{1s}}{f_{Ms}} \right \rangle_{\psi} = - \left \langle \int d^3 v \, \frac{f_{1s} \mathbb{L}_{0s} g_{1s}}{f_{Ms}} \right \rangle_{\psi}. 
\end{gather}
Therefore, in the DKES trajectory model we obtain \eqref{eq:dkes_adjoint}.

\subsection{Full trajectories}
The operator under consideration for the full model is,
\begin{multline}
    \mathbb{L}_{0s} = v_{||} \bm{b} \cdot \nabla + \bm{v}_E \cdot \nabla + \frac{(1 + \xi_s^2)x_s}{2B} \bm{v}_E \cdot \nabla B  \partder{}{x_s} \\ - \frac{1-\xi_s^2}{2B\xi_s} v_{||} \bm{b} \cdot \nabla B \partder{}{\xi_s}  +  \frac{\xi_s(1-\xi_s^2)}{2B}  \bm{v}_E \cdot \nabla B \partder{}{\xi_s}.
    \label{eq:L_0s_full}
\end{multline}

 The contribution to \eqref{eq:L_0s_adjoint} from the streaming term in \eqref{eq:L_0s_full} is identical to that in the case of the DKES trajectory model, \eqref{eq:dkes_parallel_streaming}. We next consider the contribution from the $\bm{E} \times \bm{B}$ drift term in \eqref{eq:L_0s_full}, 
\begin{align}
    \left \langle \int d^3 v \, \frac{g_{1s} \bm{v}_E \cdot \nabla f_{1s}}{f_{Ms}} \right \rangle_{\psi}
    &= - \left \langle \int d^3 v \, \frac{f_{1s} B^2 \bm{v}_E \cdot \nabla \left(g_{1s}/B^2\right)}{f_{Ms}} \right \rangle_{\psi}, 
    \label{eq:full_ve}
\end{align}
again using \eqref{eq:B_times_nabla_psi}. The contribution from the $\dot{x}_s$ term in \eqref{eq:L_0s_full} is, 
\begin{multline}
    \left \langle \int d^3 v \, \frac{g_{1s} \dot{x}_s }{f_{Ms}} \partder{f_{1s}}{x_s} \right \rangle_{\psi} = - \left \langle \int d^3 v \, \frac{f_{1s} \dot{x}_s }{f_{Ms}} \partder{g_{1s}}{x_s} \right \rangle_{\psi} \\ - \left \langle \int d^3 v \, (3+2x_s^2)(1+\xi_s^2) \frac{g_{1s}f_{1s}}{2f_{Ms}B} \bm{v}_E \cdot \nabla B \right \rangle_{\psi}. 
    \label{eq:full_x}
\end{multline}
The contribution from the mirror term in \eqref{eq:L_0s_full} is the same as in the case of the DKES trajectories model \eqref{eq:dkes_mirror}. We consider the contribution from the final term in \eqref{eq:L_0s_full}, 
\begin{multline}
    \left \langle \int d^3 v \, \frac{g_{1s} \xi_s (1- \xi_s^2) \bm{v}_E \cdot \nabla B }{2Bf_{Ms}} \partder{f_{1s}}{\xi_s} \right \rangle_{\psi} = \\ - \left \langle \int d^3 v \, \frac{f_{1s} \xi_s (1-\xi_s^2) \bm{v}_E \cdot \nabla B}{2B f_{Ms}} \partder{g_{1s}}{\xi_s} \right \rangle_{\psi}
    - \left \langle \int d^3 v \, (1-3 \xi_s^2) \bm{v}_E \cdot \nabla B \frac{f_{1s} g_{1s}}{2Bf_M} \right \rangle_{\psi}. 
    \label{eq:full_xi_2}
\end{multline}
Combining \eqref{eq:dkes_parallel_streaming}, \eqref{eq:full_ve}, \eqref{eq:full_x}, \eqref{eq:dkes_mirror}, and \eqref{eq:full_xi_2}, we obtain
\begin{multline}
    \left \langle \int d^3 v \, \frac{g_{1s} \mathbb{L}_{0s} f_{1s}}{f_{Ms}} \right \rangle_{\psi} = - \left \langle \int d^3 v \, \frac{f_{1s} \mathbb{L}_{0s} g_{1s}}{f_{Ms}} \right \rangle_{\psi} \\ + \der{\Phi}{\psi} \frac{q_s}{T_s} \left \langle \int d^3 v \, \left( \bm{v}_{\text{m}s} \cdot \nabla \psi \right) \frac{f_{1s} g_{1s}}{f_{Ms}} \right \rangle_{\psi}. 
\end{multline}
Therefore, under the full trajectory model we obtain \eqref{eq:full_adjoint}.

\section{Symmetry of the sensitivity function}
\label{app:symmetry}

In this appendix we discuss several symmetry properties of the local sensitivity function, $S_{\mathcal{R}}$, defined through \eqref{eq:magnetic_sensitivity}. The arguments that follow are similar to those in appendix C of \cite{Landreman2018}. Throughout we will assume that $B$ is stellarator symmetric and $N_P$ symmetric. We will show that this implies $N_P$ symmetry of $S_{\mathcal{R}}$. In the limit that $E_r \rightarrow 0$, then $S_{\mathcal{R}}$ also has stellarator symmetry.

\subsection{Symmetry of $S_{\mathcal{R}}$ implied by Fourier derivatives}

First we would like to show that $S_{\mathcal{R}}$ is stellarator symmetric if and only if $\partial \mathcal{R}/\partial B_{mn}^s = 0$ for all $m$ and $n$, where we express $B$ in a Fourier series,
\begin{gather}
    B = \sum_j B_{m_j n_j}^c \cos(m_j \theta - n_j \zeta) + B_{m_j n_j}^s \sin(m_j \theta - n_j \zeta). 
\end{gather}
The perturbation, $\delta B$, is decomposed similarly. We begin with the ``if'' portion of the argument. From \eqref{eq:magnetic_sensitivity} we have, 
\begin{align}
    \partder{\mathcal{R}}{B_{mn}^s} &= V'(\psi)^{-1} \int_0^{2\pi} d \theta \int_0^{2\pi} d \zeta \, \sqrt{g} S_{\mathcal{R}} \sin(m \theta - n \zeta).
    \label{eq:dRdBmns}
\end{align}
Suppose $\partial \mathcal{R}/\partial B_{mn}^s = 0$ for all $m$ and $n$. The quantity $(\sqrt{g} S_{\mathcal{R}})$ can be represented as a Fourier series,
\begin{gather}
    \left( \sqrt{g} S_{\mathcal{R}} \right) = \sum_j A_{m_j n_j}^c \cos(m_j \theta - n_j \zeta) + A_{m_j n_j}^s \sin(m_j \theta - n_j \zeta).
    \label{eq:sqrtg_SR}
\end{gather}
From \eqref{eq:dRdBmns}, we see that $A_{m n}^s = 0$ for all $m$ and $m$. 
Thus the quantity $(\sqrt{g}S_{\mathcal{R}})$ must be even under the transformation $(\theta,\zeta) \rightarrow (-\theta,-\zeta)$. We now note that $\sqrt{g}$ must be even from \eqref{eq:jacobian} under the assumption that $B$ is stellarator symmetric. Therefore $S_{\mathcal{R}}$ must be stellarator symmetric, assuming that $\sqrt{g}$ does not vanish anywhere, which must be the case for any well-defined coordinate transformation.

We continue with the ``only if" portion of the argument. Suppose $S_{\mathcal{R}}$ is stellarator symmetric. As $\sqrt{g}$ is also stellarator symmetric, $(\sqrt{g} S_{\mathcal{R}})$ can be expressed in a Fourier series as \eqref{eq:sqrtg_SR} with $A_{mn}^s =0$ for all $m$ and $n$. Thus from \eqref{eq:dRdBmns} $\partial \mathcal{R}/\partial B_{mn}^s = 0$ for all $m$ and $n$. 

We next show that if $B$ is $N_P$ symmetric, then $S_{\mathcal{R}}$ is $N_P$ symmetric if and only if $\partial \mathcal{R}/\partial B_{mn}^c = 0$ for all $n$ that are not integer multiples of $N_P$. We begin with the ``if" portion of the argument. From \eqref{eq:magnetic_sensitivity},
\begin{gather}
    \partder{\mathcal{R}}{B_{mn}^c} = V'(\psi)^{-1} \int_0^{2\pi} d \theta \int_0^{2\pi} d \zeta \, \sqrt{g} S_{\mathcal{R}} \cos(m \theta - n \zeta) 
    \label{eq:dRdBmnc}. 
\end{gather}
Suppose $\partial \mathcal{R}/\partial B_{mn}^c = 0$ for all $n$ which are not integer multiples of $N_P$. Here $(\sqrt{g} S_{\mathcal{R}})$ can be expressed in a Fourier series as \eqref{eq:sqrtg_SR} with $A_{mn}^s = 0$ for all $m$ and $n$. Inserting the Fourier series into \eqref{eq:dRdBmnc}, we find that $A_{mn}^c = 0$ for all $n$ that are not integer multiples of $N_P$. Thus $(\sqrt{g} S_{\mathcal{R}})$ must be $N_P$ symmetric. As $\sqrt{g}$ must be $N_P$ symmetric, this implies $S_{\mathcal{R}}$ possesses the same symmetry.

Next we consider the ``only if'' portion of the argument. Suppose that $S_{\mathcal{R}}$ is $N_P$ symmetric. As $\sqrt{g}$ is also $N_P$ symmetric, then $(\sqrt{g}S_{\mathcal{R}})$ can be expressed in a Fourier series as \eqref{eq:sqrtg_SR} where the sum includes $n$ that are integer multiples of $N_P$. Inserting the Fourier series into \eqref{eq:dRdBmnc}, we find that $\partial \mathcal{R}/\partial B_{mn}^c = 0$ for all $n$ that are not integer multiples of $N_P$. 

\subsection{Symmetry of Fourier derivatives}

To continue, we need to show that $\partial \mathcal{R}/\partial B_{mn}^s = 0$ for all $m$ and $n$ and $\partial \mathcal{R}/\partial B_{mn}^c = 0$ for all $n$ which are not integer multiples of $N_P$. We begin with the $N_P$ symmetry argument. We consider the symmetry of $f_{1s}$ implied by \eqref{eq:dke_model}. Under the transformation $\zeta\rightarrow \zeta + 2\pi/N_P$, we find that each of the trajectory coefficients remain unchanged, as well as the source term and collision operator. Therefore we can conclude that $f_{1s}$ is $N_P$ symmetric. We can also note that each of the $\widetilde{\mathcal{R}}$ vectors are $N_P$ symmetric, as well as $\sqrt{g}$. We consider the integrand that appears in the flux surface average in \eqref{eq:inner_product_R},
\begin{gather}
    D_s(\theta,\zeta) = \int d^3 v \, \frac{f_{1s} \widetilde{\mathcal{R}}_s^f \sqrt{g}}{f_{Ms}}.
\end{gather}
Here the superscript and subscript on $\widetilde{\mathcal{R}}$ denotes that we consider the unknowns corresponding to the distribution function of species $s$. We note that $D_s(\theta,\zeta+2\pi/N_P) = D_s(\theta,\zeta)$. The quantity $\mathcal{R}$ can be expressed in terms of $D_s$ as follows,
\begin{gather}
\mathcal{R} = \sum_s V'(\psi)^{-1}\int_0^{2\pi} d \zeta \int_0^{2\pi} d \theta \, D_s.
\label{eq:R}
\end{gather}
Next we consider the functional derivative of $\mathcal{R}$ with respect to $B$, defined as in \eqref{eq:functional_derivative}. The derivative with respect to $B_{mn}^c$ can be thus defined as,
\begin{gather}
    \partder{\mathcal{R}}{B_{mn}^c} = V'(\psi)^{-1} \int_0^{2\pi} d \zeta \int_0^{2\pi} d \theta \, \left( \sum_s \frac{\delta D_s}{\delta B} - \mathcal{R} \frac{\delta \sqrt{g}}{\delta B} \right) \cos(m \theta - n \zeta).
    \label{eq:C}
\end{gather}
As the functional derivative maintains the $N_P$ symmetry of $D_s$ and $\sqrt{g}$, the quantity in parenthesis in \eqref{eq:C} can be expressed in a Fourier series containing only $n$ that are integer multiples of $N_P$. Thus we see that the quantity $\partial \mathcal{R}/\partial B_{mn}^c = 0$ for all $n$ that are not integer multiples of $N_P$. 

Next we consider a similar argument for stellarator symmetry. We begin by considering the symmetry of $f_{1s}$ implied by \eqref{eq:dke_model} in the case $E_r = 0$. Under the transformation $(\theta,\zeta,v_{||}) \rightarrow (-\theta,-\zeta,-v_{||})$, we see that both the collisionless trajectory operator and the collision operator maintain the parity of $f_{1s}$, while the source term is odd. Therefore, $f_{1s}$ must be odd under this transformation. In this case, we can write $f_{1s}$ as \begin{gather}
    f_{1s} = f^{-}_{a,s}(x_s,\xi_s)f^+_{b,s}(\theta,\zeta) + f^{+}_{a,s}(x_s,\xi_s)f^-_{b,s}(\theta,\zeta)
    \label{eq:f_1s}
\end{gather}
where $f^-_{a,s}(x_s, -\xi_s) = -f^-_{a,s}(x_s, \xi_s)$,  $f^+_{a,s}(x_s, -\xi_s) = f^+_{a,s} (x_s, -\xi_s)$, and analogous expressions for $f_{b,s}^+$ and $f_{b,s}^-$.

We next note that each of the $\widetilde{\mathcal{R}}^f_s$ are odd under the transformation $(\theta,\zeta,v_{||}) \rightarrow (-\theta,-\zeta,-v_{||})$. As $\sqrt{g}$ is even, then we can express $\widetilde{\mathcal{R}}_s^f\sqrt{g}$ in a similar way to \eqref{eq:f_1s},
\begin{gather}
    \widetilde{\mathcal{R}}_s^f\sqrt{g}= B_{a,s}^-(x_s,\xi_s) B_{b,s}^+(\theta,\zeta)+B_{a,s}^+(x_s,\xi_s)B_{b,s}^-(\theta,\zeta). 
\end{gather}
The integrand that appears in the flux surface average becomes,
\begin{multline}
    D_s = \int d^3 v \, f_{Ms}^{-1} \bigg( f_{a,s}^-(x_s,\xi_s) B_{a,s}^-(x_s,\xi_s) f_{b,s}^+(\theta,\zeta) B_{b,s}^+(\theta,\zeta) \\ + f_{a,s}^+(x_s,\xi_s)B_{a,s}^+(x_s,\xi_s) f_{b,s}^-(\theta,\zeta)B_{b,s}^-(\theta,\zeta) \bigg).
\end{multline}
We see that $D_s$ is even with respect to the transformation $(\theta,\zeta) \rightarrow (-\theta,-\zeta)$. 
The quantity $\mathcal{R}$ can be written as in \eqref{eq:R} and the derivative with respect to a stellarator asymmetric mode is
\begin{gather}
    \partder{\mathcal{R}}{B_{mn}^s} = V'(\psi)^{-1} \int_0^{2\pi} d \zeta \int_0^{2\pi} d \theta \, \left(\sum_s \frac{\delta D_s}{\delta B} - \mathcal{R} \frac{\delta \sqrt{g}}{\delta B} \right) \sin(m \theta - n \zeta). 
\end{gather}
The functional derivative with respect to $B$ does not change the parity of $D_s$ or $\sqrt{g}$, thus we see that the quantity in parenthesis in the above equation is even with respect to the transformation $(\theta,\zeta) \rightarrow (-\theta,-\zeta)$. Therefore, $\partial \mathcal{R}/\partial B_{mn}^s = 0$ for all $m$ and $n$. A similar argument cannot be made if $E_r \neq 0$, as the inhomogeneous drive term in \eqref{eq:dke_model} no longer has definite parity. However, according to the arguments in \citep{Hirshman1986b} the transport coefficients do obey this symmetry property. 

\section{Derivatives at ambipolarity}
\label{app:ambipolar}

In this appendix we derive an expression for derivatives of moments of the distribution function at fixed ambipolarity rather than fixed $E_r$ by determining the relationship between geometry parameters, $\Omega$, and $E_r$. To begin, it is assumed that the continuous adjoint approach outlined in section \ref{sec:continuous} is used. The approach taken here is analogous to that used in appendix A of  \cite{Paul2018}, in which an additional adjoint equation is used to compute derivatives at a fixed constraint function for optimization of stellarator coil shapes. 

Consider the set of unknowns computed with SFINCS, $F$, which depends on parameters $\Omega$ and $E_r$. The total  differential of $F$ satisfies
\begin{gather}
   \mathbb{L} dF(\Omega,E_r) = \left( \partder{\mathbb{S}}{E_r}-\partder{\mathbb{L}}{E_r} F\right) dE_r + \sum_{i=1}^{N_{\Omega}}  \left( \partder{\mathbb{S}}{\Omega_i} - \partder{\mathbb{L}}{\Omega_i}F\right)  d \Omega_i,
    \label{eq:dF}
\end{gather}
which follows from \eqref{eq:linear}. Consider $J_r(F,\Omega)$, which depends on $E_r$ through $F$. The total differential of $J_r$ can be computed,
\begin{gather} dJ_r(F(\Omega,E_r),\Omega) = \sum_{i=1}^{N_{\Omega}} \left(\partder{J_r}{\Omega_i}\right)_F + \left \langle \widetilde{J_r}, dF(\Omega,E_r) \right \rangle,
\end{gather}
which can be written using \eqref{eq:dF} and the solution to \eqref{eq:J_r_adjoint},
\begin{multline}
    dJ_r(F(\Omega,E_r),\Omega) = \left \langle q^{J_r}, \left( \partder{\mathbb{S}}{E_r} - \partder{\mathbb{L}}{E_r} F \right) \right \rangle d E_r \\
    + \sum_{i=1}^{N_{\Omega}} \left(\left(\partder{J_r}{\Omega_i}\right)_F + \left \langle q^{J_r}, \left(\partder{\mathbb{S}}{\Omega_i} - \partder{\mathbb{L}}{\Omega_i} F \right) \right  \rangle  \right) d \Omega_i.
\end{multline}
By enforcing $dJ_r(F(\Omega,E_r),\Omega) = 0$, we obtain the relationship between $E_r$ and $\Omega$ at ambipolarity,
\begin{gather}
    \left( \partder{E_r}{\Omega_i} \right)_{J_r} = -\left \langle q^{J_r}, \left( \partder{\mathbb{S}}{E_r}-\partder{\mathbb{L}}{E_r} F\right) \right \rangle^{-1} \left(\left(\partder{J_r}{\Omega_i}\right)_F + \left \langle q^{J_r}, \left( \partder{\mathbb{S}}{\Omega_i} - \partder{\mathbb{L}}{\Omega_i}F \right)\right \rangle \right).
    \label{eq:dErdOmega}
\end{gather}
Consider a moment of the distribution function, $\mathcal{R}(F,\Omega)$. The derivative with respect to $\Omega_i$ at fixed ambipolarity can thus be computed,
\begin{gather}
    \left(\partder{\mathcal{R}}{\Omega_i}\right)_{J_r} = \left( \partder{\mathcal{R}}{\Omega_i} \right)_{F,E_r} + \left \langle \widetilde{\mathcal{R}}, \left(\partder{F}{\Omega_i}\right)_{J_r} \right \rangle. 
\end{gather}
The first term corresponds to the explicit dependence on $\Omega_i$, while the second contains dependence through $F$. 
Here $\left(\partial F/\partial \Omega_i\right)_{J_r}$ satisfies,
\begin{multline}
    \mathbb{L} \left( \partder{F}{\Omega_i} \right)_{J_r} = -  \left(  \partder{\mathbb{S}}{E_r} - \partder{\mathbb{L}}{E_r} F\right) \left \langle q^{J_r}, \left(\partder{\mathbb{S}}{E_r} - \partder{\mathbb{L}}{E_r}F \right) \right \rangle^{-1} \\ \times \left( \left(\partder{J_r}{\Omega_i}\right)_F + \left \langle q^{J_r}, \left(\partder{\mathbb{S}}{\Omega_i} - \partder{\mathbb{L}}{\Omega_i} F\right) \right \rangle \right) + \left(\partder{\mathbb{S}}{\Omega_i} - \partder{\mathbb{L}}{\Omega_i} F\right),
    \label{eq:dFdOmega}
\end{multline}
from \eqref{eq:dF} using \eqref{eq:dErdOmega}.
Using \eqref{eq:dFdOmega} and \eqref{eq:adjoint}, we find
\begin{multline}
    \left( \partder{\mathcal{R}}{\Omega_i} \right)_{J_r} = \left(\partder{\mathcal{R}}{\Omega_i}\right)_{F,E_r} + \left \langle q^{\mathcal{R}}, \left( \partder{\mathbb{S}}{\Omega_i}- \partder{\mathbb{L}}{\Omega_i} F \right) \right \rangle
    \\ - \left \langle q^{\mathcal{R}}, \left(\partder{\mathbb{S}}{E_r} - \partder{\mathbb{L}}{E_r} F\right) \right \rangle \frac{\left(\left(\partder{J_r}{\Omega_i}\right)_{F} + \left \langle q^{J_r}, \left(\partder{\mathbb{S}}{\Omega_i} - \partder{\mathbb{L}}{\Omega_i} F \right)\right \rangle \right)}{\left \langle q^{J_r}, \left( \partder{\mathbb{S}}{E_r} - \partder{\mathbb{L}}{E_r} F \right) \right \rangle}. 
\end{multline}
An analogous expression can be obtained using the discrete approach,
\begin{multline}
    \left( \partder{\mathcal{R}}{\Omega_i} \right)_{J_r} = \left(\partder{\mathcal{R}}{\Omega_i}\right)_{F,E_r} + \left \langle \overrightarrow{\bm{q}}^{\mathcal{R}}, \left( \partder{\overrightarrow{\bm{S}}}{\Omega_i}- \partder{\overleftrightarrow{\bm{L}}}{\Omega_i} \overrightarrow{\bm{F}} \right) \right \rangle
    \\ - \left \langle \overrightarrow{\bm{q}}^{\mathcal{R}}, \left(\partder{\overrightarrow{\bm{S}}}{E_r} - \partder{\overleftrightarrow{\bm{L}}}{E_r} \overrightarrow{\bm{F}} \right) \right \rangle \frac{\left(\left(\partder{J_r}{\Omega_i}\right)_F + \left \langle \overrightarrow{\bm{q}}^{J_r}, \left( \partder{\overrightarrow{\bm{S}}}{\Omega_i}- \partder{\overleftrightarrow{\bm{L}}}{\Omega_i} \overrightarrow{\bm{F}} \right)\right \rangle \right)}{\left \langle \overrightarrow{\bm{q}}^{J_r}, \left(\partder{\overrightarrow{\bm{S}}}{E_r} - \partder{\overleftrightarrow{\bm{L}}}{E_r} \overrightarrow{\bm{F}} \right) \right \rangle},
\end{multline}
where \eqref{eq:J_r_adjoint_discrete} has been used. 

\bibliographystyle{jpp}
\bibliography{bibliography}
\end{document}